\documentclass[preprint]{emulateapj}

\def\lsim{\raise0.3ex\hbox{$<$}\kern-0.75em{\lower0.65ex\hbox{$\sim$}}}
\def\gsim{\raise0.3ex\hbox{$>$}\kern-0.75em{\lower0.65ex\hbox{$\sim$}}}

\begin{document}

\shorttitle{Missing fractions}
\shortauthors{Mattila et al.}
\title{CORE-COLLAPSE SUPERNOVAE MISSED BY OPTICAL SURVEYS}


\author{S. Mattila\altaffilmark{1}, T. Dahlen\altaffilmark{2}, A. Efstathiou\altaffilmark{3}, E. Kankare\altaffilmark{1}, J. Melinder\altaffilmark{4}, A. Alonso-Herrero\altaffilmark{5}, M.\'A. P\'erez-Torres\altaffilmark{6}, S. Ryder\altaffilmark{7}, P. V\"ais\"anen\altaffilmark{8}, G. \"Ostlin\altaffilmark{4}}
\altaffiltext{1}{Tuorla Observatory, Department of Physics and Astronomy, University of Turku, V\"ais\"al\"antie 20, FI-21500 Piikki\"o, Finland}
\altaffiltext{2}{Space Telescope Science Institute, 3700 San Martin Drive, Baltimore, MD 21218, USA}
\altaffiltext{3}{School of Sciences, European University Cyprus, Diogenes Street, Engomi, 1516 Nicosia, Cyprus}
\altaffiltext{4}{Department of Astronomy, Oskar Klein Centre, Stockholm University, AlbaNova University Centre, 106 91, Stockholm, Sweden}
\altaffiltext{5}{Instituto de Fisica de Cantabria, CSIC-UC, Avenida de los Castros s/n, 39005 Santander, Spain}
\altaffiltext{6}{Instituto de Astrof\'{\i}sica de Andaluc\'{\i}a - CSIC, P.O. Box 3004, 18008 Granada,  Spain}
\altaffiltext{7}{Australian Astronomical Observatory, P.O. Box 296, Epping, NSW 1710, Australia}
\altaffiltext{8}{South African Astronomical Observatory, P.O. Box 9, Observatory 7935, Cape Town, South Africa}

\email{sepmat@utu.fi}



\begin{abstract}
We estimate the fraction of core-collapse supernovae (CCSNe) that remain undetected by optical
SN searches due to obscuration by large amounts of dust in their host galaxies.
This effect is especially important in luminous and ultraluminous infrared galaxies, which
are locally rare but dominate the star formation at redshifts of $z$ $\sim$ 1-2. We perform
a detailed investigation of the SN activity in the nearby luminous infrared galaxy
Arp~299 and estimate that up to 83\% of the SNe in Arp~299 and in similar
galaxies in the local Universe are missed by observations at optical wavelengths. For
rest-frame optical surveys we find
the fraction of SNe missed due to high dust extinction to increase from the average
local value of $\sim$19\% to $\sim$38\% at $z$ $\sim$ 1.2 and then stay roughly constant up to
$z$ $\sim$ 2. It is therefore crucial to take into account the effects of obscuration by dust
when determining SN rates at high redshift and when predicting the number of CCSNe detectable
by future high-$z$ surveys such as LSST, JWST, and Euclid. For a sample
of nearby CCSNe (distances 6-15 Mpc) detected during the last 12 yr, we
find a lower limit for the local CCSN rate of 1.5$^{+0.4}_{-0.3}$ $\times 10^{-4}$ yr$^{-1}$Mpc$^{-3}$,
consistent with that expected from the star formation rate. Even closer, at distances less than
$\sim$6 Mpc, we find a significant increase in the CCSN rate, indicating a local
overdensity of star formation caused by a small number of galaxies that have each hosted multiple
SNe.
\end{abstract}


\keywords{supernovae: general -- supernovae: individual(SN 2005at, SN 2010P) --- galaxies: starburst
-- galaxies: individual(\objectname{Arp 299}) -- infrared: galaxies}



\section{Introduction}
Much of the massive star formation and hence a substantial fraction of the
core-collapse supernovae (CCSNe) in the universe may be hidden behind dust. At
higher redshifts, obscured star formation in luminous (10$^{11}$$L_{\odot}$ $\leq$
$L_{{\rm IR}}$ $<$ 10$^{12}$$L_{\odot}$) and ultraluminous ($L_{{\rm IR}}$ $>$ 10$^{12}$$L_{\odot}$)
infrared (IR) galaxies (LIRGs and ULIRGs, respectively) actually dominates
over the star formation seen in the ultraviolet (UV) and optical (e.g., Le Floc'h
et al. 2005; Magnelli et al. 2009, 2011). Because of the observed concentration of star formation within the
innermost nuclear regions (e.g., Soifer et al. 2001) high spatial resolution is crucial for
detecting SNe in these environments. This can be achieved with observations
from space, with ground-based adaptive optics (AO) imaging
observations at near-IR wavelengths, or with interferometric radio imaging. High
spatial resolution searches at near-IR wavelengths have already discovered
several obscured SNe within a few hundred parsecs from LIRG nuclei
(Mattila et al. 2007; Kankare et al. 2008, 2012). Furthermore,
high spatial resolution searches at radio wavelengths have
been able to reveal SN factories within the innermost $\sim$100 pc LIRG nuclear regions
that have so far remained hidden at all other wavelengths (e.g. Lonsdale et al. 2006;
P\'erez-Torres et al. 2009; Ulvestad 2009; Romero-Ca\~nizales et al. 2011, 2012; Bondi et al. 2012; Herrero-Illana et al. 2012).

The effects of host galaxy extinction on the detectability of SNe are expected to increase with redshift
since in general, shorter rest-frame wavelengths are observed at higher
redshifts. Even more important, the fraction of the star formation hidden from optical searches in
LIRGs and ULIRGs is expected to increase rapidly toward redshift $z$ $\sim$ 1
(P\'erez-Gonz\'alez et al. 2005; Le Floc'h et al. 2005; Caputi et al. 2007; Magnelli et al. 2009, 2011).
Unless properly corrected for, errors in derived CCSN rates at $z$ $\sim$ 1 will be dominated by
these effects (Mannucci et al. 2007; Dahlen et al. 2012; Melinder et al. 2012). Recent
CCSN rate studies (e.g., Dahlen et al. 2004; Botticella et al. 2008; Bazin et al. 2009; Graur et al. 2011; Li et al. 2011a; Horiuchi et al. 2011)
have indicated that the cosmic CCSN rate might not match the massive star formation rate (SFR) even in the local Universe.
This could be caused by a population of SNe remaining undetected by the current optical searches either
because they are intrinsically faint, or dark due to large host galaxy extinctions (e.g., Horiuchi et al. 2011). Also,
SNe with lower extinctions but occurring within a few hundred parsecs of an LIRG nucleus would likely be detectable
only by observations with a high spatial resolution (e.g., Kankare et al. 2012) typically not available
for the current SN searches. More recently, Botticella et al. (2012) used a sample of 14 CCSNe
within the 11 Mpc volume to derive a robust lower limit for the local
CCSN rate. They found the volumetric CCSN rate to be consistent with that expected
from the SFR derived from far-UV luminosities and higher than expected based on H$\alpha$
luminosities. This indicates that most of the intrinsically faint and/or dark events were detected
in their local sample.

The fraction of missing SNe as a function of redshift has been studied before by
Mannucci et al. (2007). They compiled the star formation densities for different
redshifts derived from UV and IR observations.
They used these results together with their own estimates on how many SNe are
lost due to obscuration by dust in local starburst galaxies, LIRGs and ULIRGs
to derive a correction for SN rates at high redshifts. These estimates,
however, were based on a very small number of SNe detected in such galaxies by
that time (Maiolino et al. 2002; Mannucci et al. 2003). Also, at that time, very little
was known about the nature of the high redshift LIRGs and ULIRGs. The assumption
made in Mannucci et al. (2007) was that they were the same kind of systems
as in the local universe, which was not an unreasonable assumption. However, later developments,
in particular the recent results from $Spitzer$ and $Herschel$, have shown that the high-redshift U/LIRG
population is dominated by disk galaxies forming stars in the 'normal' extended (the so called
main sequence) mode rather than in compact starbursts as observed in the local U/LIRGs
(e.g., Elbaz et al. 2011).

In this investigation we use a somewhat similar approach to that of Mannucci et al. (2007) to
estimate the corrections needed in order to account for CCSNe remaining undetected
by optical surveys both locally and as a function of redshift. Our corrections
consist of two parts. First, we estimate the fraction of CCSNe in normal galaxies
with substantially higher host galaxy extinctions than predicted by simple models
for the smooth dust distribution and the resulting inclination effects. Thereafter,
we estimate the fraction of CCSNe missed
in local U/LIRGs. For this we make use of the rich SN population of one of the
nearest LIRGs, Arp 299. Assuming that the SNe with the highest host galaxy extinctions are
missed by the optical searches and not compensated for by the standard extinction
corrections, we can derive the fraction of SNe that remain missing and estimate the
corrections needed to be applied when deriving CCSN rates. We then use this information
together with the latest knowledge of the nature and evolution of high-$z$ U/LIRGs
to calculate the fraction of CCSNe missed as a function of redshift. We assume
H$_{0}$ = 70 km s$^{-1}$ Mpc$^{-1}$, $\Omega_{\Lambda}$ = 0.7, and $\Omega_{M}$ = 0.3 throughout
the paper.

\section{Supernova budget of nearby galaxies}

\subsection{The Nearby Supernova Sample}
In order to investigate the completeness of the current local optical SN searches at different
distances we selected all the SNe discovered by the end of 2011 and after the beginning of the year
2000 from the Asiago SN catalog (Barbon et al. 1999) with an identified host galaxy and v$_{rec}$ $<$ 1500 km s$^{-1}$
(corresponding to distances less than $\sim$20 Mpc). We then adopted their radial velocities corrected for the
peculiar velocities due to Virgo infall from the HyperLeda database (Paturel et al. 2003) and converted to distances.
We ended-up with a sample of 100 objects including also a few events without a spectroscopic classification,
Type Ia's and luminous blue variable (LBV) outbursts originally classified as SNe.
We then investigated the volumetric rate of these transients within different distances over the 12 yr
period. A significantly higher rate was found at the smallest distances of $\sim$4-6 Mpc indicative of
the local overdensity of star formation (see the discussion below). The rate was found to stay roughly constant between
$\sim$12 and $\sim$15 Mpc and then show a significant decline. The drop of the rate at distances greater than $\sim$15 Mpc
could be a result of the SN searches starting to miss a significant fraction of both the intrinsically faint and heavily
dust-obscured events after this point (see also Botticella et al. 2012 and Horiuchi et al. 2011).

\begin{table*}
\caption{Volume-limited Sample of SNe Closer Than 15 Mpc Discovered in 2000-2011}
\centering
\begin{tabular}{lccccccccccc}
\hline
\hline
SN & Host galaxy & Velocity & Distance & Method & Type & A$_{\rm V}$ host (reference) & A$_{\rm V}$ & M$_{\rm peak}$ & Incl. & Botticella & log L$_{\rm IR}$\\
   &             & (km s$^{-1}$) & (Mpc)&        &      &                & MW         &           & ($^{\circ}$) & & (L$_{\odot}$)\\
\tableline
2000db & NGC 3949 & 1020 & 14.6 & Kinematic & IIP & ... & 0.07 & ... & 57 & No & 9.93 \\
2001ig & NGC 7424 & 754 & 10.8 & Kinematic & IIb & $<$0.06$^{8}$ (Silverman et al. 2009) & 0.03 & -17.4$^{8}$ & 59 & No & - \\
2002ap & NGC 628 & 686 & 9.3$^{1}$ & Mean & Ic & 0.04 (Takada-Hidai et al. 2002) & 0.24 & -17.7$^{2}$ & 35 & Yes & 9.89\\
2002bu & NGC 4242 & 737 & 10.5 & Kinematic & 08S-like & ... & 0.04 & ... &52 & Yes & ...\\
2002hh & NGC 6946 & 318 & 5.7$^{2}$ & Mean & IIP & 4.1 (Pozzo et al. 2006) & 1.13 & -18.4$^{2}$ & 31 & Yes & 9.94\\
2003J & NGC 4157 & 1011 & 14.4 & Kinematic & IIP & ... & 0.07 & ... & 90 & No & 10.19\\
2003gd & NGC 628 & 686 & 9.3$^{1}$ & Mean & IIP & 0.2 (Smartt et al. 2009) & 0.23 & -16.6$^{2}$ & 35 & Yes & 9.89\\
2003ie & NGC 4051 & 917 & 13.1 & Kinematic & IIP & ... & 0.04 & ... & 30 & No & 9.90\\
2003jg & NGC 2997 & 914 & 13.1 & Kinematic & Ib/c & 3.7 (N. Elias-Rosa et al. in prep.) & 0.36 & -17.8$^{8}$ & 32 & No & ...\\
2004am & M 82 & 487 & 3.3$^{3}$ & Cepheid & IIP & $\sim$5 (Mattila et al. 2012) & 0.53 & -15.4$^{12}$ & 79 & Yes & 10.66\\
2004dj & NGC 2403 & 370 & 3.3$^{3}$ & Cepheid & IIP & 0.4 (Smartt et al. 2009) & 0.13 & -16.6$^{2}$ & 60 & Yes & 9.21\\
2004et & NGC 6946 & 318 & 5.7$^{2}$ & Mean & IIP & 0.2 (Smartt et al. 2009) & 1.13 & -17.9$^{2}$ & 31 & Yes & 9.94\\
2005ae & ESO 209-G009 & 862 & 12.3 & Kinematic & IIb & ... & 0.86 & - & 90 & No & 9.85\\
2005af & NGC 4945 & 376 & 3.8$^{4}$ & TRGB & IIP & $\sim$0 (Pereyra et al. 2006) & 0.61 & -15.7$^{2}$ & 90 & Yes & 10.45\\
2005at & NGC 6744 & 618 & 8.8 & Kinematic & Ic & 2.3$\pm$0.3 (Kankare et al. in prep.) & 0.14 & -16.1$^{2}$ & 54 & Yes & 9.89\\
2005ay & NGC 3938 & 1017 & 14.5 & Kinematic & IIP & ... & 0.07 & ... & 14 & No & 9.92\\
2005cs & M 51 & 702 & 10.0 & Kinematic & IIP & 0.3 (Smartt et al. 2009) & 0.12 & -15.9$^{2}$ & 30 & Yes & ...\\
2006my & NGC 4651 & 912 & 13.0 & Kinematic & II & ... & 0.09 & ... & 50 & No & 9.58\\
2007gr & NGC 1058 & 634 & 9.3$^{5}$ & Cepheid & Ic & 0.09 (Hunter et al. 2009) & 0.21 & -17.5$^{2}$ & 20 & Yes & ...\\
2007it & NGC 5530 & 1046 & 14.9 & Kinematic & II & ... & 0.39 & - & 67 & No & ...\\
2008S & NGC 6946 & 318 & 5.7$^{2}$ & Mean & 08S-like & ... & 1.12 & ... & 31 & No & 9.94\\
N300-OT & NGC 300 & -38 & 1.9$^{6}$ & Cepheid & 08S-like & ... & 0.04 & ... & 40 & No & 8.35\\
2008ax & NGC 4490 & 797 & 11.4 & Kinematic & IIb & 1.5 (Chornock et al. 2011) & 0.07 & -18.5$^{2}$ & 47 & Yes & 10.28 \\
2008bk & NGC 7793 & 60 & 3.4$^{7}$ & Cepheid & IIP & $\sim$0 (Van Dyk et al. 2012) & 0.07 & -15.2$^{2}$ & 53 & Yes & 8.92 \\
2008iz & M 82 & 487 & 3.3$^{3}$ & Cepheid & RSN & $\lesssim$10 (Mattila et al. 2012)  & 0.53 & ... & 79 & No & 10.66\\
2008jb & ESO 302-G014 & 636 & 9.1 & Kinematic & IIP & 0.19 (Prieto et al. 2012) & 0.03 & -15.3$^{9}$ & 74 & No & ...\\
2009N & NGC 4487 & 1026 & 14.7 & Kinematic & IIP & ... & 0.07 & ... & 46 & No & ...\\
2009dd & NGC 4088 & 989 & 14.1 & Kinematic & II & ... & 0.07 & ... & 71 & No & 10.29\\
2009hd & NGC 3627 & 788 & 9.4$^{3}$ & Cepheid & II & 3.7 (Elias-Rosa et al. 2011) & 0.11 & -17.7$^{2}$ & 57 & Yes & 10.33\\
2009ib & NGC 1559 & 1004 & 14.3 & Kinematic & IIP & ... & 0.10 & ... & 60 & No & 10.31 \\
2009ls & NGC 3423 & 1032 & 14.7 & Kinematic & II & ... & 0.10 & ... & 32 & No & ...\\
2010br & NGC 4051 & 917 & 13.1 & Kinematic & Ib/c & ... & 0.04 & ... & 30 & No & 9.90\\
2010dn & NGC 3184 & 766 & 10.9 & Kinematic & 08S-like & ... & 0.06 & ... & 24 & No & 9.67\\
2011dh & M 51 & 702 & 10.0 & Kinematic & IIb & $<$0.15 (Arcavi et al. 2011) & 0.12 & -17.0$^{10}$ & 30 & No & ...\\
2011ja & NGC 4945 & 376 & 3.8$^{4}$ & TRGB & IIP & $>3$ (Monard et al. 2011) & 0.59 & -17.5$^{11}$ & 90 & No & 10.45\\
2011jm & NGC 4809 & 979 & 14.0 & Kinematic & Ic & ... & 0.11 & ... & 90 & No & ...\\
\tableline
\end{tabular}
\tablecomments{
$^{1}$Hendry et al. 2005; $^{2}$Botticella et al. 2009; $^{3}$Freedman et al. 2001; $^{4}$Karachentsev et al. 2007; 
$^{5}$NGC 1058, belongs to a group of nearby galaxies of which NGC 925 is also a member - Silbermann et al. 1996;
$^{6}$Gieren et al. 2005; $^{7}$Pietrzy\'nski et al. 2010; $^{8}$Horiuchi et al. 2011; $^{9}$Prieto et al. 2012;
$^{10}$Arcavi et al. 2011; $^{11}$Monard et al. 2011; $^{12}$Mattila et al. (2012).}
\end{table*}

In the following analysis we therefore only consider the 49 events with Virgo infall corrected recession velocities
of less than 1050 km s$^{-1}$ (or distances less than 15 Mpc). We excluded any Type Ia SNe (SNe 2001el, 2005df, 2005ew, 2006E, 2006mq,
2008ge, 2010ae and 2011fe) and LBV outbursts (SNe 2000ch, 2002kg and 2010da; Wagner et al. 2004; Pastorello et al. 2010; Smith
et al. 2011; Maund et al. 2006). The initial sample also includes one event, SN 2008eh, without a spectroscopic
classification. Based on its discovery magnitude and some light curve information Horiuchi et al. (2011) assumed that it was an
intrinsically faint CCSN. However, without any certain information on the nature of this event, we decide to exclude it from
our final sample. Two of the remaining events with estimated distances larger than 15 Mpc were also excluded from our analysis.
SN 2003hn occurred in NGC 1448 which also hosted the well observed normal Type Ia SN 2001el.
The optical and near-IR photometry of SN 2001el yielded a distance of 17.9 Mpc for NGC 1448 (Krisciunas et al. 2003; Mattila
et al. 2005b). The host galaxy of SN 2004gk with a negative (Virgo infall corrected) recession velocity has a Tully-Fisher based
distance of $\sim$20 Mpc (Solanes et al. 2002). We are therefore left with a total of 35 events that are listed in Table 1.

We also initially included SN 2008S-like events as a part of the analysis. We have therefore added the NGC 300-2008OT
in Table 1, which is an event similar to SN 2008S and missing from the Asiago catalog. In total, our initial sample includes
four such events: SNe 2002bu, 2008S, 2010dn, and NGC 300-2008OT. An origin in an explosion of an extreme asymptotic
giant branch star as an electron capture SN is favored for the SN 2008S-like events by a number
of authors (e.g., Botticella et al. 2009; Pumo et al. 2009). However, several others found a non-explosive
outburst of a massive star the most plausible origin (e.g., Bond et al. 2009; Berger et al. 2009; Smith et al.
2009; Kashi et al. 2010). Recent optical observations of the SN 2008S site (Szczygiel et al. 2012) have already
ruled out the presence of a massive evolved progenitor star indicating that either it did not survive the 2008
event or has now returned to its dust enshrouded state. Kochanek (2011) predicts that the shock powering the
current IR luminosity from the site of SN 2008S should destroy the dust, eventually allowing a direct
confirmation if the progenitor of SN 2008S has disappeared. With the origin of the SN 2008S-like
events still an open question, we decided to exclude them from our current analysis. We note that the
SN 2008S-like events were also excluded by Botticella et al. (2012) from their CCSN rate analysis.

Our sample also includes the recently reported SN 2008jb that remained undetected by all the
pointed SN searches despite a distance of only $\sim$~9 Mpc (Prieto et al. 2012). 
It occurred in a low luminosity host galaxy, ESO 302-G014, similar to the Magellanic Clouds and not
included in the target lists of the pointed SN searches. The fraction of CCSNe occurring in such
low luminosity/metallicity galaxies is very low; $<$2\% according to Young et al.
(2008). Therefore, such events missed by optical pointed SN searches locally are not
likely to contribute significantly to the fraction of CCSNe missed by optical
SN searches in general. 

The adopted host galaxy distances are mostly the kinematic distances obtained from the Virgo infall
corrected recession velocities. However, for several of the closest
galaxies, more accurate distances were available from Cepheid observations or from the tip of the
red giant branch (TRGB) method which were adopted instead. In a few cases the mean distance
from several methods was adopted from the literature. The adopted distances are listed in Table 1 with
the relevant references given in the notes to the table. In Table 1, we also list the IR luminosities of
the host galaxies (if available from Sanders et al. 2003) scaled to our adopted distances.

The volumetric CCSN rates obtained using the events from Table 1 are listed in Table 2.
We adopt the small number statistical uncertainties from Gehrels (1986) and use these throughout
this study. Excluding the SN 2008S-like events there are a total of eight CCSNe within 6 Mpc, and the CCSN rate
within 6 Mpc becomes 7.4 $^{+3.7}_{-2.6}$ $\times$ 10$^{-4}$ yr$^{-1}$ Mpc$^{-3}$. This is
significantly higher than the volumetric CCSN rate reported by Botticella et al. (2012), and
that expected from the SFR extrapolated from high-$z$ and even the locally
normalized Horiuchi SFR. However, such an elevated CCSN rate is consistent with a local overdensity of
star formation observed within $\sim$10 Mpc (e.g., Karachentsev et al. 2004). Furthermore, 
according to Heckman (1998) only four galaxies (M 82, NGC 253, M 83 and NGC 4945) are responsible for 25\% of
all the high-mass ($\geq$ 8 M$_{\odot}$) star formation within 10 Mpc distance. It is interesting that all of these
four galaxies are actually closer than 6 Mpc. Of the total of eight CCSNe within 6 Mpc, SNe 2004am and 2008iz occurred
in M 82, SNe 2005af and 2011ja in NGC 4945, and SNe 2002hh and 2004et in NGC 6946.
So, there are only five host galaxies within 6 Mpc that are responsible for all eight events
over the last 12 yr within this volume. Also, it is possible that the SFR within 6 Mpc is actually
concentrated in more dusty systems such as the prototypical starburst galaxy M 82, and, therefore,
the missing fraction of SNe within this volume might be higher than representative
for the local volume.

In order to avoid the effects of overdensity, we calculated the volumetric CCSN rates
after excluding the volume within 6 Mpc. These rates are listed in Table 2 for different
outer limits of the volume. We note that between 10 and 15 Mpc this volumetric CCSN rate
stays constant within statistical uncertainties indicating that within 15 Mpc we
are not missing a significant number of events. The volumetric CCSN rate at 6-15 Mpc is
1.5 $^{+0.4}_{-0.3}$ $\times$ 10$^{-4}$ yr$^{-1}$ Mpc$^{-3}$, which is similar
to the volumetric CCSN rate recently derived by Botticella et al. (2012)
for a similar sample of CCSNe within 11 Mpc (the 14 events used by Botticella et al. are indicated
in Column 11 of Table 1). This value is also consistent with the expected CCSN rate from the local
star formation density (see Horiuchi et al. 2011). 

\subsection{Host galaxy extinctions}
For all the SNe within 12 Mpc except SN 2005at host galaxy extinctions were available in the literature (for
references see Column 7 of Table 1). We note that these extinctions were estimated for each SN, not for each
galaxy as a whole. Therefore, here we concentrate on the CCSNe within 12 Mpc which is
almost identical to the distance limit in the sample of Botticella et al. (2012) when accounting for the difference
in the value of $H_{0}$ adopted in the two studies. The Galactic extinctions were adopted from Schlegel et al. (1998).
For SN 2005at which is a Type Ic event resembling SN 1994I (see Schmidt \& Salvo 2005) at $\sim$9 Mpc,
very little information was available in the literature. In E. Kankare et al. (in preparation) we present
the available $UBVRI$ photometry of SN 2005at, together with the optical spectrum of Schmidt \& Salvo (2005),
covering a wide wavelength range from $\sim$3300 to $\sim$10200 \AA. We use these data to estimate a
host galaxy extinction of $A_{\rm V}$ = 2.3 $\pm$ 0.3 for SN 2005at by comparison with well observed
normal Type Ic events.

In Table 3, we compare the observed extinction properties of our CCSN sample within 12 Mpc with
the predictions from our Monte Carlo simulations (E. Kankare et al. in preparation). For this comparison we include
the 18 CCSNe from Table 1, excluding the SN 2008S-like events whose origin is still an open question.
These simulations follow the recipe in Riello \& Patat (2005), assuming a homogeneous dust distribution for the model host galaxy 
with the bulge-to-total ($B/T$) ratio = 0.0 (100\% of the SNe located in the disk) and an optical depth through a
simulated face-on galaxy at a zero radius of $\tau_{V}$(0) = 2.5, consistent with many statistical studies (e.g., Kankare et al. 2009),
to derive a distribution of expected $A_V$ values. For the
purpose of comparing observed and predicted extinction values, we divided our sample into three inclination bins.
The host galaxy inclinations were adopted from the HyperLeda database.

First, we compare the observed fraction of CCSNe in different inclination bins with the predictions
from the simulations (see Columns 7 and 11 in Table 3). The relative distribution of SNe in galaxies with
inclinations of 0-30$^{\circ}$ and 30-60$^{\circ}$ appears to be as expected whereas there is an apparent lack
of SNe in host galaxies with an inclinations higher than 60$^{\circ}$. Although half of the SNe would be expected
in galaxies with inclination of 61-90$^{\circ}$, less than 30\% were discovered in such galaxies. In the edge-on
bin $\sim$10\% of the SNe have a simulated extinction of $A_{\rm V}>5$. Therefore, we do expect a fraction of
the SNe to be missed in normal galaxies with the highest inclinations. However, the expected median
extinction of objects in the edge-on bin is $\sim$0.6 in $A_{\rm V}$, which means that the optical
SN searches should have been sensitive to most of the SNe even in the edge-on galaxies within the
12 Mpc volume. Therefore, the apparent lack of events in the edge-on bin appears to be more likely due
to the selection effect of the SN searches avoiding these galaxies rather than being a consequence of
the host galaxy extinction. The fact that there is an apparent lack of CCSNe with edge-on host galaxies
indicates that our SN sample within 12 Mpc (and the one used by Botticella et al. 2012)
is likely not complete. Therefore, the local volumetric CCSN rate could be higher than estimated above, which
indicates that the local overdensity of star formation could also affect the results within the 6-15 Mpc
distance.

\begin{table}
\caption{Volumetric CCSN Rates Using the Events from Table 1}
\centering
\begin{tabular}{lcccccr}
\hline
\hline
Distance & &\multicolumn{5}{c}{~~~~~~~~~~~~~~~SN Rate (10$^{-4}$ yr$^{-1}$ Mpc$^{-3}$)~~~~~~~~~~} \\
(Mpc) & &\multicolumn{2}{c}{~~~~CCSNe+08S-like~~~~~~~} & & \multicolumn{2}{c}{~~~~~~~CCSNe~~~~~~~} \\
\tableline\tableline
5 & &11.1 (7) & ... & &9.5 (6) & ... (0)\\
6 & &9.2 (10) & ... & &7.4 (8) & ... (0)\\
7 & &5.8 (10) & 3.1 (2) & &4.6 (8) & 0 (0)\\
8 & &3.9 (10) & 1.3 (2) & &3.1 (8) & 0 (0)\\
9 & &3.0 (11) & 1.2 (3) & &2.5 (9) & 0.39 (1)\\
10 & &3.6 (18) & 2.5 (10) & &3.2 (16) & 2.0 (8)\\
11 & &3.1 (21) & 2.3 (13) & &2.5 (17) & 1.6 (9)\\
12 & &2.5 (22) & 1.8 (14) & &2.1 (18) & 1.3 (10)\\
13 & &2.2 (24) & 1.6 (14) & &1.8 (20) & 1.2 (12)\\
14 & &2.0 (28) & 1.6 (20) & &1.7 (24) & 1.3 (16)\\
15 & &2.1 (36) & 1.8 (28) & &1.9 (32) & 1.5 (24)\\
\tableline
\end{tabular}
\tablecomments{The rates including the entire volume within the given distance
are given in Columns 2 and 4, and the rates excluding the volume within 6 Mpc are given in Columns 3 and 5.
The number of events included is given in parenthesis.}
\end{table}

\begin{table*}
\caption{Comparison between the Observed and Predicted Host Galaxy Extinctions.}
\centering
\begin{tabular}{lcccccccccccc}
\hline
\hline
Inclination & &\multicolumn{6}{c}{Observed}                & &\multicolumn{4}{c}{Predicted} \\ \cline{2-8} \cline{10-13}
            && $<A_{\rm V}$$>$ & Median &SEM  & $\sigma$ & SNe  & Fraction & &$<A_{\rm V}$$>$ & Median  & $\sigma$ & Fraction\\
\tableline\tableline
0 $\leq$ $i$ $\leq$ 30 & & 0.18 & 0.15 & 0.06 & 0.11    & 3 & 17\% && 0.37 & 0.15 & 0.52 & 13.5\%\\
30 $<$ $i$ $\leq$ 60 && 1.25 & 0.40 & 0.50 & 1.59    & 10 & 55\% && 0.52 & 0.20 & 0.73& 36.4\%\\
60 $<$ $i$ $\leq$ 90$^{a}$ && 1.06 & 0.19 & 0.97  & 1.68 & 5 & 28\% && 1.89 & 0.57 & 3.58& 50.1\%\\\hline
Sample 1&&&&&&&&&\\
0 $\leq$ $i$ $\leq$ 60 && 1.00 & 0.20 & 0.40 & 1.45 & 13 & 72\% && 0.48 & 0.18 & 0.68& 49.9\%\\
Sample 2$^{b}$&&&&&&&&&\\
0 $\leq$ $i$ $\leq$ 60 && 0.48 & 0.20 & 0.22  & 0.74 & 11 & 72\% && 0.48 & 0.18 & 0.68& 49.9\%\\
\tableline
\end{tabular}
\tablecomments{
The predicted distributions follow that of Riello \& Patat (2005)
for B/T = 0.0 and assuming $\tau_{V}$(0) = 2.5 adopted to match the observations.\\
$^{a}$Excluding SNe 2004am and 2008iz (both being within the nuclear regions of the
prototypical starburst galaxy M82) and therefore based on three events.\\
$^{b}$Excluding SNe 2002hh and 2009hd with the highest host galaxy extinctions from the sample.}
\end{table*}

\begin{figure}
\epsscale{1.0}
\plotone{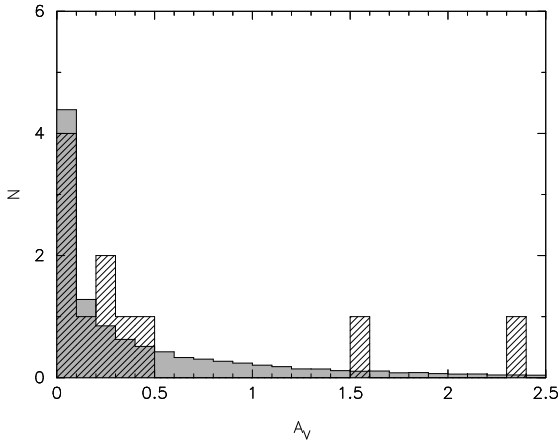}
\caption{Comparison between the observed (hatched area) and the predicted (gray area)
distribution of extinctions ($A_{\rm V}$) for CCSN host galaxy inclinations between 0$^{\circ}$ and 60$^{\circ}$.
The CCSNe closer than 12 Mpc have been included and the two outliers (SNe 2002hh and 2009hd)
have been excluded. The two events with the highest extinctions in the sample are SNe 2008ax
and 2005at. The predicted distribution (with the absolute level scaled to match the number of
observed SNe) follows that of Riello \& Patat (2005) for $B/T$ = 0.0 but assuming $\tau_{V}$(0) = 2.5.}
\label{figrate}
\end{figure}

Second, we compare the observed host galaxy extinctions in different
inclination bins with the predictions (see Table 3). In the face-on (0-30$^{\circ}$) bin the observed average
and median extinctions of $A_{\rm V}$ = 0.18 and 0.15, respectively, are quite low, similar to the predicted values
although any quantitative comparison between the extinctions is difficult because of the small
number of events. However, in the intermediate inclination (30-60$^{\circ}$) bin we have 10 SNe,
allowing for a more meaningful comparison. The average and median extinctions for this sample are
1.25 and 0.40 in $A_{\rm V}$, which are both significantly higher than their predicted values of 0.52 and 0.20,
respectively, indicating that the smooth distribution of dust used in the simulations does not agree
well with the observations. In the edge-on (60-90$^{\circ}$) sample we have five events, two of them (SNe 2004am and 2008iz)
within the nuclear regions of the prototypical nearby starburst galaxy M 82. SN 2008iz was discovered at radio
wavelengths with no reported detection in the optical indicating a very high extinction being
within the nuclear regions of M 82 (e.g., Brunthaler et al. 2010). Based on a detection in the near-IR $K$-band,
Mattila et al. (2012) estimated the likely extinction toward SN 2008iz to be not more than $A_{\rm V}$ $\sim$ 10.
SN 2004am occurred coincident
with the obscured super star cluster M82-L, and has an estimated host galaxy extinction of A$_{\rm V}$ $\sim$ 5
(Mattila et al. 2012). Excluding these two events the average extinction for the remaining three
events occurring in normal spiral galaxies is $\sim$1.1, which is lower than predicted but of course highly
uncertain due to the very small number of events included.

\subsection{Intrinsically Faint Supernovae}
Using our adopted distances and extinctions we calculated optical absolute peak magnitudes for the events
within a distance of 12 Mpc. These are mostly based on the SN peak magnitudes available from Horiuchi et al.
(2011) and Botticella et al. (2012). These absolute SN peak magnitudes are very approximate especially
for a couple of the most recent events for which we adopted their discovery magnitudes (for references see the note to
Table 1). We find that 3 of the 17 CCSNe within 12 Mpc are fainter than $M$ = -15.5 but none of them fainter
than $M$ = -15.0. We note that the absolute SN magnitudes considered by Horiuchi et al. (2011) were not corrected
for host galaxy extinctions. After applying the correction we find that from their intrinsically faint CCSNe
only SN 2004am has an absolute peak magnitude of $\sim$ -15 (see Column 9 in Table 1). These faint SNe are more likely
to be missed in a search over a larger volume compared to the restriction $<12$Mpc for our sample and could therefore
lead to underestimates of the CCSN rate. The absolute peak magnitudes listed in Table 1 (excluding SN 2008iz with
no reported optical detection) therefore suggest that a correction factor approximately $\sim$1.2 may actually be needed
to account for the intrinsically faint CCSNe in a survey that is not deep enough for their detection. However, such
intrinsically faint events are best accounted for by including a realistic distribution of the absolute peak magnitudes
when deriving the CCSN rate results. For example, the effects of including a large (30\%) fraction of faint ($M >$ -15)
CCSNe was studied by Melinder et al. (2012). As a result they found an increase of $\sim$30\% in their CCSN rates at
$z$ $\sim$ 0.4 and $z$ $\sim$ 0.7.

\subsection{Missing SNe in Normal Galaxies}
We can now proceed to estimate the missing fraction of SNe due to high dust extinctions
in normal galaxies ($L_{\rm IR}$ $<$ 10$^{11}$ $L_{\odot}$) using all the available information we have gathered.
A correction for the dust extinction along the lines of Riello \& Patat (2005) can be used to compensate for the CCSNe
missed due to inclination effects. However, this should be combined with a fraction of SNe with significantly higher
extinctions than predicted by these models when considering a realistic distribution of extinctions for CCSNe in normal
galaxies. We concentrate on the sample of 13 CCSNe with host galaxy inclinations of 0-60$^{\circ}$ (see Table 3).
The host galaxies of these events have IR luminosities ranging between $\sim$10$^{9}$ and $\sim$3 $\times$ 10$^{10}$
$L_{\odot}$ (see Column 12, Table 1). The observed average extinction calculated
for the 0-60$^{\circ}$ bin of $<A_{\rm V}>$ = 1.00 (sample 1 in Table 3) is significantly higher than
the predicted value, $<A_{\rm V}>$ = 0.48 (although the median extinctions are similar). However, if
excluding the two events with the highest host galaxy extinctions (SNe 2002hh and 2009hd with A$_{\rm V}$ = 3.7
and 4.1, respectively) the observed average extinction becomes $<$A$_{\rm V}$$>$ = 0.48 (sample 2 in Table 3), which is
identical to the predicted value (the median value does not change). Also the standard deviations of the observed and
predicted distributions are now very similar (0.74 versus 0.68). 

In Figure 1, the observed and predicted extinction distributions are compared after excluding the two outliers
with $A_{\rm V}$ $\gsim$ 3.7. We note that only about 0.3$\%$ of the events in the predicted distribution suffer from
extinctions higher than $A_{\rm V}$ = 3.7. This indicates that the simple model with a homogeneous dust distribution
used for the simulations does not produce a realistic distribution of line-of-sight extinctions for SNe in spiral
galaxies. CCSNe are observed to be associated with star-forming regions in their host galaxies
(e.g., Anderson et al. 2012) and therefore higher extinction values local to the SNe can be expected. However, we
also note that the number of SNe in our current sample is rather small making any detailed comparison less reliable.

The two outlier events with host galaxy extinctions of $A_{\rm V}$ = 3.7 - 4.1 (or $A_{\rm B}$ = 4.9 - 5.5)
would be the ones most likely missed by optical SN searches in more distant galaxies. For example, a typical Type II-P event
with absolute peak magnitude $M(B)$ = -17.0 at a redshift of 0.5 would have an apparent peak magnitude
of $m(R)$ $\sim$ 30 if the host galaxy extinction is $A_{\rm B}$ = 5. This is several magnitudes below the limits of the current
high-$z$ SN searches (e.g., Dahlen et al. 2012, Melinder et al. 2012). At higher redshifts the effects of the
extinction would be even more severe for an optical SN search since shorter rest-frame wavelengths are observed.
However, we note that SN searches are often optimized for detecting SNe at rest-frame $B$- and $V$-bands, since the SN spectral
energy distribution (SED) peaks in this range. For example, searches aiming at a redshift around 0.5 (e.g., Bazin et al. 2009;
Melinder et al. 2012)
have been typically observing in $R$- and $I$-bands and searches reaching for $z$ $\sim$ 1 (e.g., Dahlen et al. 2012) in the z-band.
The ongoing SN search program (e.g., Rodney et al. 2012) as a part of the CANDELS project (e.g., Grogin et al. 2011)
is aiming to detect SNe at $z$ $\sim$1.5-2 and observes in
the near-IR $J$- and $H$-bands. Therefore, we now use the sample of 13 CCSNe within the 0$^{\circ}$-60$^{\circ}$ bin and the two outlier
events with the highest host galaxy extinctions to estimate the fraction of CCSNe likely missed by rest-frame optical surveys
in normal galaxies to be 15$^{+21}_{-10}$\%.

\section{Supernova budget of LIRGs and ULIRGs}
We next consider the fraction of missing SNe in galaxies with $L_{\rm IR}$ $>$ 10$^{11}$ $L_{\odot}$
i.e. LIRGs and ULIRGs. The interacting system Arp 299 (=IC 694 + NGC 3690) is one of the most nearby
examples of an LIRG at a luminosity distance of 46.7 Mpc. Its IR luminosity $L_{\rm IR}$ =
L[8-1000$\mu$m] of 7.3 $\times$ 10$^{11}$ $L_{\odot}$ (adopted from Sanders et al. 2003 and
scaled to the assumed distance) indicates a very high CCSN rate of $\sim$2 yr$^{-1}$ which is
one of the highest expected in local galaxies. Arp 299 has been the target of several SN
searches and over the last two decades a total of seven SNe have been discovered at optical or
near-IR wavelengths within its circumnuclear regions, $\sim$1-4 kpc from the main galaxy nuclei
A and B1 (see Figure 2). It is therefore well suited for investigating the fraction of SNe missed
by optical observations in one of the best observed local LIRGs. In Section 4, we make use of these
estimates to also extrapolate the results to higher redshifts.

\begin{figure}[t]
\epsscale{1.0}
\includegraphics[scale=0.7,angle=-90]{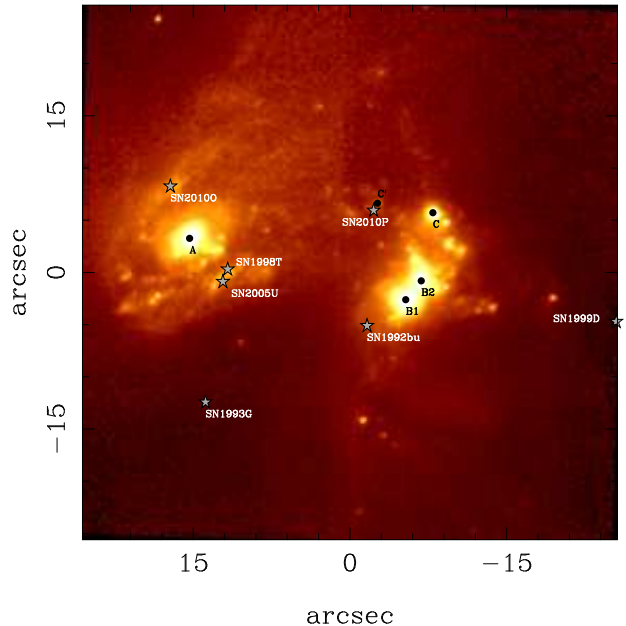}
\caption{$HST$/NICMOS F164N image (from Alonso-Herrero et al. 2000) of Arp 299 shown with a square-root scaling
to emphasize the extent of the diffuse emission in the circumnuclear regions. This image traces the CCSN
activity via the [Fe II] 1.644~$\mu$m line. The positions of the main nuclei A, B1 and B2, sources C and C$^{\prime}$,
and the SNe discovered at optical or IR wavelengths are indicated.}
\end{figure}

\subsection{Observed SNe in Arp 299}
During the last two decades Arp 299 has been the target of a number of
professional and amateur SN searches (see Table 4 for summary). At optical
wavelengths, Arp 299 has been the target of at least four professional SN
searches. In the early 1990s it was monitored by the Leuschner Observatory
SN Search that was the predecessor of the Lick Observatory SN search (LOSS), the
Beijing Observatory SN search, the Richmond et al. (1998) search and finally the LOSS.
Arp 299 was also included in the 'optimal' galaxy sample of the LOSS
with their SN search data between 1998 March and 2008 December used for
the SN rate calculation (Leaman et al. 2011; Li et al. 2011a, 2011b). SN 1993G
was discovered by the Leuschner Observatory SN Search, SN 1998T by the
Beijing Observatory SN search and SN 1999D by the LOSS. LOSS also detected
SNe 1998T and 2005U, although they were originally discovered elsewhere. The observed $B-V$
color of SN 1993G indicated that the SN was virtually unreddened (Tsvetkov 1994).
SN 1998T was spectroscopically classified by Li et al. (1998)
reporting strong P-Cygni profiles of He I lines between 510 and 900 nm. Therefore,
the extinction toward SN 1998T was also likely low. In the case of SN 1999D
the typing spectrum featured a very blue continuum (Jha et al. 1999), strongly
suggesting a low extinction. For these three events we therefore assume
that the host galaxy extinction was close to zero.

\begin{table}
\caption{SN Searches in Arp 299}
\centering
\begin{tabular}{lccc}
\hline
\hline
Telescope & Band & Period & SNe\\
\tableline
Leuschner$^{a}$ & vis & Before 1998         & 1993G\\
Richmond & VIS & 1998 Dec-1991 Jun & ...\\
BAOSS$^{b}$& VIS &              & SN 1998T\\
LOSS$^{c}$ & VIS & 1998-ongoing & 1999D (1998T, 2005U)\\\hline
IRTF & NIR & 1992 Mar-1993 Dec & 1992bu\\
WIRO & NIR & 1993 Feb-1993 Dec &...\\
TNG & NIR & 1999 Oct-2001 Oct & ...\\
USNO & NIR & 2001 Feb-2004 May & ...\\
WHT & NIR & 2002 Jan-2005 Jan & 2005U\\
NOT & NIR & 2005 Mar-ongoing & 2010P (2010O)\\
Gemini-N & NIR & 2008 Apr-ongoing & (2010O, 2010P)\\\hline
\tableline
\end{tabular}
\tablecomments{$^{a}$Leuschner Observatory Supernova Search.
$^{b}$Beijing Astronomical Observatory Supernova Survey.
$^{c}$Lick Observatory Supernova Search.}
\end{table}

Van Buren et al. (1994) conducted a $K$-band survey for SNe in starburst
galaxies at the NASA 3.0 m Infrared Telescope Facility (IRTF). As a part of
their search Arp 299 was observed  in at least four epochs between 1992 March
and 1993 December, yielding the discovery of SN~1992bu in their $K$-band
images. The SN is located at a projected distance of $\sim$1.0 kpc
from the nucleus B1 of the galaxy. SN 1992bu has no spectroscopic classification,
and based on just the $K$-band photometry, estimating the extinction is difficult.
However, as a part of their analysis Anderson et al. (2011) noted that SN
1992bu falls on a bright star-forming region and has a small galactocentric distance.
Hence, they find it consistent with being a stripped envelope core-collapse
event.

A more recent near-IR SN search targeting nearby starburst galaxies was
carried out in the $K'$-band by Grossan et al. (1999). The observations
were performed with the 2.3~m telescope at the Wyoming IR Observatory (WIRO). As part of the
search, five observations of Arp 299 were taken between 1993 February and December.
Arp 299 was also targeted for a near-IR $K$-band SN search carried out with
the 3.6 m Telescopio Nazionale Galileo (TNG) between 1999 October and 2001 October including
nine observations (Mannucci et al. 2003). Another near-IR $K$-band SN search observing
Arp 299 was carried out using the 1.55 m U.S. Naval Observatory telescope at
Flagstaff (e.g., Dudley et al. 2008). During the search Arp 299 was observed
in a total of 19 epochs between 2001 February and 2004 May (C. Dudley, private
communication). However, no confirmed SNe were reported in Arp 299 as a result of these
three searches.

\begin{table}
\caption{The Optical/Near-IR SNe of Arp 299}
\centering
\begin{tabular}{lccccc}
\hline
\hline
SN & SN Type & Distance (kpc) & $A_{\rm V}$ & Band & Discovery Date\\
\tableline
1992bu & ... & 1.0 (B1) & ... & NIR & 1992 Mar 9\\
1993G & IIL & 3.6 (A) & $\sim$0 & VIS & 1993 Mar 5\\
1998T & II & 1.1 (A) & $\sim$0 & VIS & 1998 Mar 3 \\
1999D & Ib & 4.5 (B1) & $\sim$0  & VIS & 1999 Jan 17\\
2005U & IIb & 1.2 (A) & $\sim$0 & NIR & 2005 Jan 30\\
2010O & Ib & 1.2 (A) & $\sim$2 & VIS & 2010 Jan 24\\
2010P & Ib/IIb & 0.2 (C$^{\prime}$) & $\sim$5 & NIR & 2010 Jan 18\\
\tableline
\end{tabular}
\tablecomments{The offsets are from the nearest nucleus (given in brackets) using their
coordinates at 8.46 GHz from Romero-Ca\~nizales et al. (2011). For identifications of the
supernovae and the different nuclei see Figure 2.}
\end{table}

Arp 299 was also monitored in near-IR $K$-band for a total of eight epochs
between 2002 January and 2005 January as a part of the nuclear SN search campaign
with the William Herschel Telescope resulting in the discovery of SN 2005U
(Mattila et al. 2004, 2005a). The near-IR color of SN 2005U indicated only
a modest extinction. The SN was spectroscopically classified as a Type-II
event, probably within a few weeks of the explosion (Modjaz et al.
2005). A further spectrum of SN 2005U was obtained by Leonard \& Cenko (2005)
showing it to resemble the spectrum of the Type IIb SN 1993J over a month past
its explosion with no evidence of substantial
host galaxy extinction (D. C. Leonard, private communication; M. Modjaz et al.
in preparation).

More recently, we have been monitoring Arp 299 for SNe in the $K$-band using the 2.6~m
Nordic Optical Telescope (NOT) every few months for a total of $\sim$20 epochs,
resulting in the discovery of SN 2010P (Mattila \& Kankare 2010) and the
detection of SN 2010O, which was also discovered at optical wavelengths (Newton et al. 2010).
SN 2010O was spectroscopically classified as a type Ib/c event
around maximum light and a host galaxy extinction of $A_{\rm V}$ $\sim$ 1.9 was
estimated from prominent Na I D absorption lines from the host galaxy (Mattila
et al. 2010). Between 2008 and 2011 we have also been monitoring Arp 299 using
high spatial resolution observations with laser guide star adaptive optics on the
Gemini North telescope as part of a sample of nearby LIRGs (see Kankare et al. 2008, 2012).
Our most recent $K$-band observations from the NOT and the Gemini-N telescope
were obtained in 2012 March and 2012 February, respectively.

Recently, Anderson et al. (2011) analyzed the circumnuclear SN population in Arp 299
finding a relatively high fraction of stripped
envelope events (Types Ib and IIb) relative to other Type II SNe. They
suggested that this excess could be explained by the young age of circumnuclear
star formation in Arp 299 such that we would now be witnessing the explosions
of the most massive stars formed. Alternatively, the excess of stripped
envelope events might be explained by a top-heavy initial mass function (IMF)
favoring the formation of the most massive stars. 

\begin{figure}[t]
\includegraphics[scale=0.3,angle=-90]{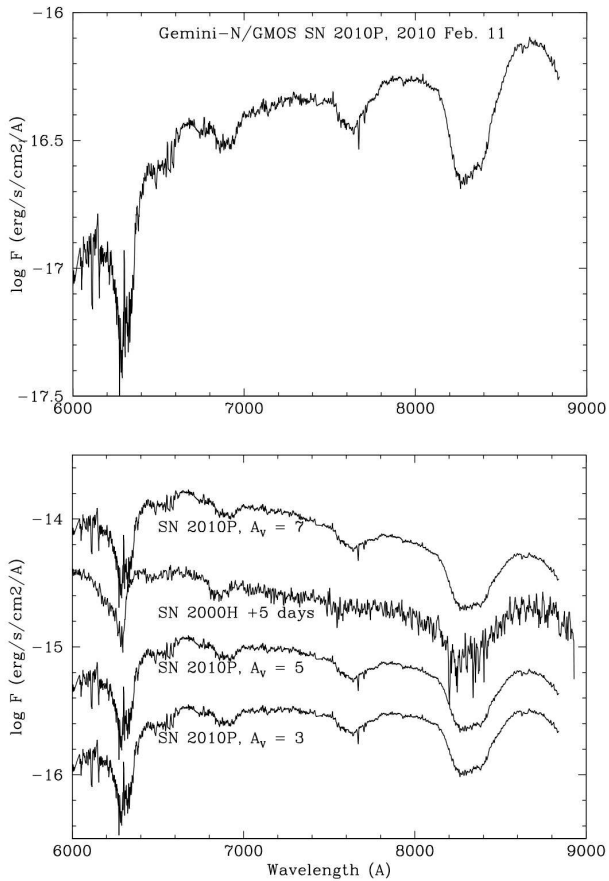}
\caption{Gemini-N/GMOS spectrum of SN 2010P within the nuclear regions
of Arp 299. Top: the observed spectrum before dereddening. Bottom:
the spectrum after dereddening by $A_{\rm V}$ = 3, 5, and 7 compared
with the Type IIb/Ib SN 2000H (corrected for the galactic reddening according to
Schlegel et al. (1998)).}
\end{figure}

\subsubsection{The highly obscured SN 2010P}
SN 2010P was discovered  via image-subtraction techniques in near-IR images
obtained using NOTCam on January 18.2 and 23.1 UT. The approximate magnitudes of SN 2010P on 2010
January 23.1 were $m_{I}$ = 18.3, $m_{J}$ = 16.8, $m_{H}$ = 16.2, $m_{K}$ = 15.9 (Mattila \& Kankare 2010).
Nothing is visible at this position on NOTCam $K$-band images taken on 2009 November 27.2 (limiting mag
17.5). We note that SN 2010P was also detectable by image subtraction in our $I$-band image obtained
using StanCam on January 23.1 on the NOT. However, the SN was not detected in our $R$-band images
from the same instrument. A radio follow-up using MERLIN between 2010 January 29 and February 1 at 4.99 GHz
resulted in a non-detection for SN 2010P (Beswick et al. 2010). The absolute magnitude and colors of
2010P are consistent with the CCSN template light curves from Mattila \& Meikle (2001) with a likely
extinction of $A_{\rm V}$ $\sim$ 5.

In order to allow spectroscopic typing and a more reliable extinction determination we obtained long-slit
spectroscopy of SN 2010P with the Gemini Multi-Object Spectrograph (GMOS; Hook et al. 2004) attached to
the Gemini-N Telescope as a part of the program GN-2010A-Q-40 (PI: S. Ryder). Four exposures of 900~sec
each were taken on the night of 2010 February 11 UT using the R400 grating and a $0\farcs5$ wide slit to
deliver a resolution of $R\sim1900$. The data were reduced and combined using V1.10 of the {\em gemini} package
within IRAF\footnote{IRAF is distributed by the National Optical Astronomy Observatories, which
are operated by the Association of Universities for Research in Astronomy, Inc., under cooperative agreement
with the National Science Foundation.}.

The spectrum of SN 2010P has a red continuum and very little signal at wavelengths shorter than 6000 \AA
(see Figure 3), indicative of a high extinction. Despite significant reddening, the spectrum was found to be
consistent with an H-deficient CCSN. Comparison with a library of SN spectra using both the "GELATO"
code (Harutyunyan et al. 2008) and the "SuperNova IDentification" code (Blondin \& Tonry 2007) yielded good
matches to several Type Ib and Type IIb SNe between 1 and 3 weeks after maximum light (the spectrum
of SN 2010P was dereddened by 5 mag of extinction in $A_{\rm V}$ before using the ``Gelato'' code). In Figure 3 we illustrate this by
comparing the spectrum of SN 2010P dereddened by $A_{\rm V}$ = 3, 5, and 7, with that of the Type IIb/Ib 
event SN 2000H at 5 days past maximum from Branch et al. (2002). For this the reddening law of
Cardelli et al. (1989) was assumed with a value of $R_{\rm V}$ = 3.1. Following Elmhamdi et al. (2006) we assume the
host galaxy extinction toward SN 2000H to be small. A very high amount
of extinction toward SN 2010P is also consistent with its near-IR light curves, colors, and absolute magnitude. 

SN 2010P has a projected distance of only $\sim$200 pc from the source C$^{\prime}$ but $\sim$1.3 kpc from the source C
(see Figure 2). Although the region C$^{\prime}$ is a strong radio source, it is much less prominent at IR wavelengths
with only $\sim$10\% of source C at 2.2$\mu$m increasing to $\sim$1/3 at 18 $\mu$m
(Charmandaris et al. 2002). Based on the optical depth of the 9.7 $\mu$m absorption
feature, Alonso-Herrero et al. (2009) estimated an extinction of $A_{\rm V}$ = 12
toward source C$^{\prime}$, which is similar to their estimated extinction for nucleus B1.
Having only $\sim$200 pc projected distance from the source C$^{\prime}$ and a substantial host galaxy extinction,
we consider SN 2010P as a genuine nuclear SN in the following analysis. Other
SNe with high host galaxy extinctions previously detected in LIRGs include SNe 2001db ($A_{\rm V}$ $\sim$ 5.5;
Maiolino et al. 2002), 2004ip ($A_{\rm V}$ $\geq$ 5; Mattila et al. 2007), 2008cs ($A_{\rm V}$ $\sim$ 16; Kankare et
al. 2008), and 2011hi ($A_{\rm V}$ = 5-7; Kankare et al. 2012; Romero-Ca\~nizales et al. 2012), all discovered in the
near-IR $K$-band similar to SN 2010P.

\begin{figure*}[t]
\epsscale{1.0}
\plotone{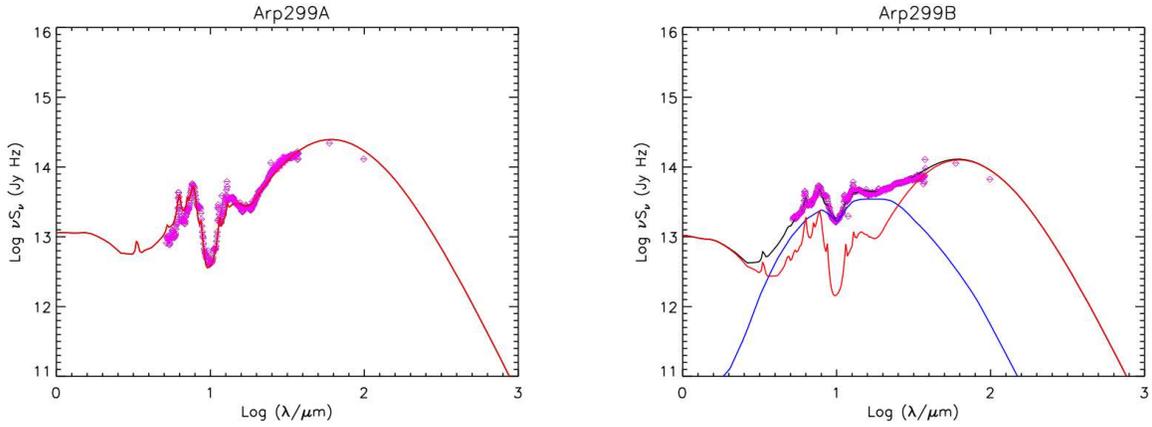}
\caption{Starburst (red line) and AGN (blue line) model fits to the mid-IR SED of
the nuclei A and B1+B2 of Arp 299. For nucleus A, no AGN component is required whereas
for nucleus B1+B2, there is a 20\% contribution by an AGN. The starburst model parameters
are given in Table 6.}
\end{figure*}

\subsection{Optical/near-IR CCSN rate of Arp 299}

Including only the SNe discovered from 1998 onwards when the LOSS and a number
of near-IR SN searches monitoring Arp 299 were active, we have a total of five SNe
discovered during a period of $\sim$14 yr (1998 March - 2012 March). The projected distances
of these SNe from the nearest nuclei are listed in Table 5. For all except SN 2010P these
range between $\sim$1 and $\sim$5 kpc.

Despite the numerous efforts at optical and near-IR wavelengths to detect SNe within
the innermost nuclear regions of Arp 299 during the last two decades, only one (SN 2010P)
has been detected close to one of the nuclei of Arp 299. This is consistent with the high
dust extinctions therein, which makes the SNe too dim to be detectable. In order to estimate
the actual fraction of SNe missed in Arp 299 due to dust extinction, we need to estimate
the fraction of star formation and thus CCSNe in the nuclear and circumnuclear regions.

From the observations we know that at least five SNe occurred in Arp 299 during the 14 yr
period. Using these five SNe we can now estimate a lower limit for the optical/near-IR CCSN rate
in Arp 299 over the period of 14 yr of 0.36 $^{+0.24}_{-0.15}$ yr$^{-1}$. Of the
five SNe all except SN 2010P were detected by the optical SN searches (although SN 2005U
was originally discovered in the near-IR) and had a relatively low extinction.
SN 2010P was marginally detected in our $I$-band images from the NOT (after its initial
discovery in the near-IR) and was not detectable in $R$-band images from the same telescope.
It had a high extinction of $A_{\rm V}$ $\sim$ 5 and would therefore most likely not be detected in
the current optical SN searches. In fact, shortly before our discovery report on SN 2010P,
the discovery of SN 2010O was reported by Newton et al. (2010)
in the course of the Puckett Observatory Supernova Search. They did not detect SN 2010P in their
unfiltered CCD images. Therefore, we assume that four out of the five SNe discovered in Arp 299
during the 14 yr period were optically detectable. This yields a lower limit for the rate of
optically detectable circumnuclear SNe in Arp 299 of 0.29 $^{+0.23}_{-0.14}$ yr$^{-1}$. 

\subsection{Predicted CCSN Rate of Arp 299}
After having determined the observed SN rate in Arp 299, we now turn to estimating the predicted rate
based on the SFR in order to calculate the fraction of SNe hidden by dust.
The CCSN rate within the highly obscured innermost nuclear regions of Arp 299
can be estimated either via infrared or radio luminosity. Radio luminosity based
estimates for the nuclear SN rate of Arp 299 from VLA measurements have been
presented by Neff et al. (2004) who estimated (scaled to the distance of 46.7 Mpc
assumed here) 0.65-1.3 yr$^{-1}$, 0.13-0.26 yr$^{-1}$, and 0.07-0.13 and 0.04-0.08 yr$^{-1}$
for the nuclei A and B1, and sources C and C$^{\prime}$ respectively (see Figure 2 for the identifications of
the different nuclei). Alonso-Herrero et al. (2000) estimated
0.8 yr$^{-1}$ and 0.15 yr$^{-1}$ for nuclei A and B, respectively (again scaled
to 46.7 Mpc). More, recently, Romero-Ca\~nizales et al. (2011) estimated a lower limit of
0.28 $^{+0.27}_{-0.15}$ yr$^{-1}$
for the CCSN rate of nucleus B1 based on archival VLA observations over a period of 11 yr,
which we will adopt as the more direct estimate to use in this study.
P\'erez-Torres et al. (2009) and Bondi et al. (2012) made use of extremely high spatial resolution
European VLBI network (EVN) radio observations of the innermost nuclear regions of Arp 299-A.
They find clear evidence for at least two new RSNe in their observations separated by 2 yr,
therefore implying a lower limit for the CCSN rate in nucleus A of $\sim$0.8 yr$^{-1}$.

We can also estimate the CCSN rates indirectly from the galaxy IR luminosity. Charmandaris et al.
(2002) estimated that IC 694 (nucleus A), NGC 3690 (nucleus B1+B2), and sources C+C$^{\prime}$ emit
approximately 39\%, 20\%, and 10\% of the total IR luminosity of Arp 299, respectively,
with the remaining $\sim$31\% originating from the circumnuclear regions (see Column 2, Table 7).
The rather large contribution of the circumnuclear regions to the total luminosity is also supported
by the H$\alpha$ observations by Garc\'ia-Mar\'in et al. (2006) showing significant amounts
of star formation in the spiral arms of IC 694 and also to the west of nucleus B1 in NGC 3690.
Adopting the galaxy IR luminosity of 7.3 $\times$ 10$^{11}$ $L_{\odot}$ and an empirical relation
between the IR luminosity and CCSN rate ($r_{\rm SN}$ = 2.7 $\times$ 10$^{-12}$ $\times$
$L_{\rm IR}$/$L_{\odot}$ yr$^{-1}$) from Mattila \& Meikle (2001), we have a total CCSN rate
of $\sim$2.0 yr$^{-1}$ for the entire system (see also the discussion in Romero-Ca\~nizales et
al. 2011). However, this IR luminosity based CCSN rate estimate could have a significant
uncertainty because the empirical relation is based on only three nearby starburst galaxies
NGC 253, M 82, and NGC 4038/9 with IR luminosities substantially lower than of Arp 299.

\subsection{Modeling the SEDs of Nuclei A and B1+B2}
As a more accurate approach we estimate the CCSN rates for the Arp 299 nuclei by
modeling their SEDs. This approach has the advantage
that it can take into account the effects of the starburst age as well
as the possible contribution of an active galactic nucleus (AGN) to the IR luminosities of
the nuclei. For this purpose we have used low resolution mid-IR $Spitzer$/IRS (SL+LL setting;
5-38$\mu$m range) spectra covering $\sim$10.4'' $\times$ 10.4'' rectangular regions 
(see the lower panel of Figure 5 in Alonso-Herrero et al. 2009) centered
on the nuclei A and B1+B2. These spectra were obtained on 2004 April 15 and have already
been reported in Alonso-Herrero et al. (2009). In addition, we included IRAS 12, 25, 60
and 100$\mu$m fluxes from Sanders et al. (2003) assigned to nuclei A and B1+B2 according
to the estimated contributions from Charmandaris et al. (2002). In order to obtain a reasonable
match between the $Spitzer$ spectra and the IRAS 12 and 25$\mu$m fluxes we multiplied both the
$Spitzer$/IRS spectra by 1.3. This value is well within the uncertainties expected in the
estimates of Charmandaris et al. (2002) for the fractions of IR luminosities arising from
the different nuclei of Arp 299 and the calibration errors of the $IRAS$ and $Spitzer$/IRS data.

For modeling the SED of Arp 299 we use a grid
of AGN torus models that have been computed with the method of Efstathiou \& Rowan-Robinson
(1995) and a grid of starburst models that have been computed with the method of Efstathiou et al. (2000).
For the AGN torus models we use the tapered disk models computed with the method of Efstathiou
\& Rowan-Robinson (1995) and described in more detail in A. Efstathiou et al. (in preparation).
These models considered a distribution of grain species and sizes, multiple scattering
and a density distribution that followed $r^{-1}$ where $r$ is the distance from the central
source. The models assumed a smooth distribution of dust, so they are a good approximation of
the density distribution in the torus if the mean distance between clouds is small compared
with the size of the torus, but they have been quite successful in fitting the SEDs of AGNs
even in cases where mid-infrared spectroscopy is available
(e.g., A. Efstathiou et al. 2012 in preparation). In this grid of models we consider
four discrete values for the equatorial 1000\AA~ optical depth (500, 750, 1000, 1250), three
values for the ratio of outer to inner disk radii (20, 60, 100) and three values for the
opening angle of the disk (30$^{\circ}$, 45$^{\circ}$ and 60$^{\circ}$). The spectra are computed for
inclinations which are equally spaced in the range 0-$\pi/2$.

Efstathiou et al. (2000) presented a starburst model that combined
the stellar population synthesis model of Bruzual \& Charlot (2003), a detailed radiative transfer
that included the effect of small grains and polycyclic aromatic hydrocarbons (PAHs), and a simple evolutionary scheme for the
molecular clouds that constitute the starburst. The model predicts the SEDs
of starburst galaxies from the ultraviolet to the millimeter as a function of
the age of the starburst and the initial optical depth of the molecular clouds. In this paper
we use a sequence of models that have been computed with an updated dust model (Efstathiou
\& Siebenmorgen 2009). We assume an exponentially declining SFR with an e-folding
time of 20 Myr. The choice of 20 Myr is supported by fits to the far-IR color-color diagrams
(e.g., Efstathiou et al 2000), to the SEDs of ULIRGs (e.g., Farrah et al. 2003), and to the Spoon
diagram (Rowan-Robinson \& Efstathiou 2009). Our model fits to the $Spitzer$/IRS (SL+LL) spectra are shown in Figure 4. For 
nucleus A, no AGN component is required, whereas for nucleus B1+B2 we find a 20\% contribution by an AGN
to the total IR (8-1000$\mu$m) luminosity. This agrees well with the recent findings of Alonso-Herrero
et al. (2012). The best-fit dusty torus parameters we find for the AGN are $\tau$(1000\AA) = 500,
$r_{outer}$ / $r_{inner}$ = 20, an opening angle of 60$^{\circ}$ and an inclination of 45$^{\circ}$. The resulting
starburst parameters are listed in Table 6.

The CCSN rates were then estimated in the following way.
The stellar population synthesis model of Bruzual \& Charlot (2003) makes a prediction of the
CCSN rate SNR(t) at a time $t$ after star formation in an instantaneous burst. The
starburst model of Efstathiou et al. (2000) predicts the spectrum
of this instantaneous burst at time $t$ and assumes a star formation history for the starburst.
It is therefore possible to calculate self-consistently the CCSN rate at different stages
in the evolution of a starburst by convolving the star formation history with SNR(t).
This results in CCSN rates of 0.76 and 0.33 yr$^{-1}$ for nuclei A and B1+B2, respectively.

\begin{table}
\caption{The Starburst Parameters Obtained from the SED Fits for Arp 299}
\centering
\begin{tabular}{llllll}
\hline
\hline
Region & Age   & $\tau_{\rm V}$ & $L_{\rm IR}$ ($L_{\odot}$)& $<$SFR$>$ & SNR \\
\tableline
A & 45 Myr          & 75    & 2.45$\times$10$^{11}$ & 90 $M_{\odot}$yr$^{-1}$& 0.76 yr$^{-1}$\\
B1+B2~~ & 55 Myr~~     & 100~~   & 1.19$\times$10$^{11}$~~ & 56 $M_{\odot}$yr$^{-1}$~~& 0.33 yr$^{-1}$\\
\tableline
\end{tabular}
\tablecomments{The IR luminosities of the starburst component are listed in Column 4.
The total IR luminosity of nucleus B is 1.49 $\times$ 10$^{11}$ $L_{\odot}$ including
a 20\% contribution from an AGN. An exponentially declining SFR
with an e-folding time of 20 Myr has been assumed. The SFRs given in Column 5
have been averaged over the duration of the starburst. The corresponding CCSN rates
(SNRs) are given in Column 6.}
\end{table}

\subsection{The Missing SNe in Arp 299}
The IR luminosities for the different components of Arp 299 are listed in Table 7 (Column 2) adopting their
fractional contributions from Charmandaris et al. (2002) and a total IR luminosity for the system
of 7.3 $\times$ 10$^{11}$ $L_{\odot}$. We compared our SED fit based model IR luminosities (see
Table 6) of nuclei A and B1+B2 with the 'observed' ones obtained as described above.
The model luminosity for nucleus A of 2.45 $\times$ 10$^{11}$ $L_{\odot}$ is slightly lower than
the observed $L_{\rm IR}$ = 2.85 $\times$ 10$^{11}$ $L_{\odot}$. However, we note that the simple gray body
models of Charmandaris et al. (2002) also gave lower IR luminosities for the Arp 299 nuclei
than the observed values. The total (AGN+starburst) model IR luminosity for nucleus B1+B2 of
1.49 $\times$ 10$^{11}$ $L_{\odot}$ is almost identical to the observed value of 1.46 $\times$
10$^{11}$ $L_{\odot}$.

For the sources C+C$^{\prime}$ we estimate the CCSN rates from the
observed IR luminosity using the empirical relation of Mattila \& Meikle (2001). This yields a
CCSN rate of 0.20 yr$^{-1}$ which is also consistent with the radio based estimate of
0.16 $\pm$ 0.05 yr$^{-1}$. We note that
the same approach would yield CCSN rate estimates of 0.77 and 0.39 yr$^{-1}$ for nuclei A and B1+B2
very similar to the IR SED modeling based values (note that a slightly higher value for nucleus B1+B2
is as expected due to the 20\% AGN contribution in B1). For the total nuclear CCSN rate we combine the
results from the IR SED modeling of nucleus A and B1+B2, with the IR luminosity derived rate for
sources C+C$^{\prime}$ to obtain a predicted nuclear CCSN rate of 1.29 yr$^{-1}$. 

The results from Charmandaris et al. (2002) suggest that 31\% of the total IR luminosity originates
outside the nuclei A, B1+B2, and C+C$^{\prime}$. Furthermore, Alonso-Herrero et al. (2009) also found evidence
from their $Spitzer$/IRS spectral mapping for significant circumnuclear PAH and [Ne II] emission extending
outside the main nuclei in Arp 299. Adopting the empirical relation between the IR luminosity
and the CCSN rate, this corresponds to 0.61 yr$^{-1}$ for the circumnuclear regions. We note that the use
of the CCSN rate estimate (for unobscured Type II+Ib/c SNe in normal galaxies) in units of the galaxy far-IR
luminosity from Cappellaro et al. (1999) would yield a very similar circumnuclear CCSN rate estimate.
Furthermore, Garc\'ia-Mar\'in et al. (2006) determined an SFR of 43 $M_{\odot}$ yr$^{-1}$ for the entire Arp 299
system (with the nucleus B1 excluded) based on their estimate for the extinction corrected H${\alpha}$
luminosity. Garc\'ia-Mar\'in et al. (2006) found most of the circumnuclear H II regions in Arp 299 to only suffer from
modest extinctions of typically less than $A_{\rm V}$ $\sim$ 1. Assuming a Salpeter IMF between 0.1 and 125 $M_{\odot}$
and CCSN progenitor masses between 8 and 50 $M_{\odot}$ (e.g., see the discussion in Melinder et al. 2012 and
Dahlen et al. 2012) this corresponds to a CCSN rate of $\sim$0.30 yr$^{-1}$ for Arp 299.
Having the H$\alpha$ emission from Arp 299 arising mostly outside the heavily obscured nuclear regions A,
B1+B2, and C+C$^{\prime}$ (in contrast to the IR luminosity), we can consider this as a robust lower limit for the
circumnuclear CCSN rate and adopt the IR luminosity based value of 0.61 yr$^{-1}$ as an upper limit. 
Combining these circumnuclear CCSN rate estimates with the nuclear CCSN rate yields therefore a total CCSN rate of
1.59-1.90 yr$^{-1}$ for Arp 299. In Table 7, our predicted
CCSN rates are compared to the observed rates based on radio, optical+NIR, and optical searches. These
are broadly consistent with each other and together with the data on the optical SN discoveries in the
circumnuclear regions can be used to estimate the missing fraction of SNe in Arp 299.

\begin{table*}
\caption{The SN Budget of Arp 299.}
\centering
\begin{tabular}{lcccccccccc}
\hline
\hline
Region  &  &~~~~~~~~$L_{\rm IR}$~~~~~~~~&& \multicolumn{3}{c}{~~~~~~Predicted SNR (yr$^{-1}$)~~~~~~}&& \multicolumn{2}{c}{~~~~~~Observed SNR (yr$^{-1}$)~~~~~~}\\ \cline{4-7} \cline{9-10}   
        & &($\times$10$^{11}$$L_{\odot}$)& & IR & Radio & $H_{\alpha}$ && Optical+NIR & Optical\\
\tableline\tableline
A     && 2.85 && 0.76 &$>$0.8 &...&&...&...\\
B1+B2 && 1.46 && 0.33 &$>$0.28 $^{+0.27}_{-0.15}$ &...&&...&...\\
C+C$^{\prime}$  && 0.73 && 0.20 &$\sim$0.16 $\pm$ 0.05&...&&$>$0.07$^{+0.17}_{-0.06}$&...\\
Circumnuclear && 2.26 && 0.61 &...& 0.30 &&$>$0.29$^{+0.22}_{-0.14}$ & $>$0.29$^{+0.22}_{-0.14}$\\\hline
Total         && 7.3 && 1.90  &$>$1.2&0.30 && $>$0.36 & $>$0.29 \\\hline
\tableline
\end{tabular}
\tablecomments{The predicted IR based CCSN rates for nuclei A and B1+B2 were obtained by radiative
transfer modeling of their IR SEDs, whereas the rates for sources C+C$^{\prime}$ and the circumnuclear
regions were obtained adopting the fractions of the IR luminosity arising from the different components
from Charmandaris et al. (2002) and using the empirical relation between the IR luminosity and CCSN rate
from Mattila \& Meikle (2001). These yield a total CCSN rate of 1.90 yr$^{-1}$ for Arp 299. The predicted
radio based CCSN rates for nuclei A and B1+B2, and sources C+C$^{\prime}$ were adopted from Bondi et al. (2012),
Romero-Ca\~nizales et al. (2011), and Neff et al. (2004), respectively.}
\end{table*}

We are now ready to estimate the missing fractions. Within the nuclear regions, no SNe have been
detected by optical observations (and only one, SN 2010P, by near-IR observations) and we therefore
assume that 100\% of the SNe are missed by optical searches. For the circumnuclear regions, we estimated
a lower limit for the 'optical' CCSN rate of
0.29 $^{+0.23}_{-0.14}$ yr$^{-1}$, while the predicted rate is 0.30-0.61 yr$^{-1}$, suggesting a missing fraction
of up to 37 $^{+38}_{-37}$ \% in this region. Compared to the total predicted
CCSN rate in Arp 299 of 1.59-1.91 yr$^{-1}$, we estimate a missing fraction of up to
83 $^{+9}_{-15}$ \%, taking into account both statistical errors and the uncertainty in the total CCSN rate.
We adopt these values as the missing fractions of CCSNe in local LIRGs (see Table 8).

\begin{table}
\caption{The Missing SN Fraction in Different Types of Galaxies}
\centering
\begin{tabular}{ll}
\hline
\hline
Galaxy & Missing Fraction \\
\tableline\tableline
Normal (observed) & 15 $^{+21}_{-10}$ \% \\
Normal (corrected) & 19 $^{+24}_{-12}$ \% \\
U/LIRG (circumnuclear) & 37 $^{+38}_{-37}$ \% \\
U/LIRG (nuclear) & 100 \% \\
U/LIRG (total) & 83 $^{+9}_{-15}$\%\\
\tableline
\end{tabular}
\tablecomments{The assumed missing SN fractions in normal galaxies, in the nuclear and circumnuclear regions
of U/LIRGs, and their combination (total) used to calculate the missing SN fraction as a function of redshift
(see Table 10) according to the assumptions given in Table 9. The missing fractions in normal galaxies are given
both as observed and after correcting for the contribution of unobscured SFR as traced by the UV light.}
\end{table}

\section{Missing supernovae as a function of redshift}
From observations of local infrared luminous galaxies, we know that a large fraction of the SNe
exploding in these galaxies are invisible to optical searches. For the local LIRG Arp 299,
we have estimated above that $\sim$83\% of the SNe have probably been missed by optical
observations. The local ULIRG Arp 220 has, similar to Arp 299, been monitored for SNe by
several programs (e.g., Richmond et al. 1998; Leaman et al. 2011) without confirmed SN detection
at optical or near-IR wavelengths despite the large CCSN rate inferred by interferometric
radio observations (e.g., Lonsdale et al. 2006). Therefore, close to
100\% of the SNe in such local ULIRGs are probably missed by optical searches. With an increased
fraction of the total star formation occurring in LIRGs and ULIRGs at higher redshifts to at least
$z$ $\sim$ 2 (Le Floc'h et al. 2005; Magnelli et al. 2009, 2011), it is expected that the fraction of
SNe missed should also increase with look-back time.

\begin{table}
\caption{Models for the Missing SN Fraction}
\centering
\begin{tabular}{lcccc}
\hline
\hline
Model    &~~~~~~~~Normal~~~~~~~~& \multicolumn{3}{c}{U/LIRGs}\\ \cline{3-5}
  &       & Local & Starburst & Non-starburst \\
\tableline\tableline
Nominal & 19\% & 83\% & 83\% & 37\% \\ 
Low & 7\% & 83\% & 83\% & 19\% \\
High & 43\% & 83\% & 83\% & 83\%\\
\tableline
\end{tabular}
\tablecomments{The missing SN fractions in normal galaxies, local U/LIRGs and
in U/LIRGs at $z$ $>$ 0 which are starbursting and non-starbursting.}
\end{table}

Using the missing fractions derived from local LIRGs and ULIRGs is, however,
complicated by the fact that they do not have the same properties as their
non-local counterparts. In fact, high redshift ULIRGs seem to be more similar
to local LIRGs than to Arp 220, showing more extended regions of star formation
(e.g., Rujopakarn et al 2011). While local ULIRGs have often a single nucleus and are in
an advanced stage of merging (Veilleux et al. 2002), Kartaltepe et al. (2012)
showed that high redshift U/LIRGs show a wide range of morphological
types including mergers, interactions, pure spheroids and non-interacting disks.
Furthermore, Kartaltepe et al. found that in total $\sim$43\% of their high-redshift
($z>1.5$) sample of LIRGs and ULIRGs are starbursting, defined as
having a specific SFR a factor 3 higher than the star-forming main sequence
(Elbaz et al. 2011).

Alonso-Herrero et al. (2009) have demonstrated a good match between the PAH
features of Arp 299 in their integrated mid-IR spectrum of the whole system
and in the spectra of
high-$z$ ULIRGs. For the integrated spectrum of Arp 299 they used a large extraction
aperture including all the nuclei and also a significant amount of circumnuclear
star formation in regions between IC 694 and NGC 3690. The level of obscuration in
Arp 299 appears to be similar to that in M 82, sub millimeter galaxies. and the high-$z$ ULIRGs in
the sample of Farrah et al. (2008), but it is lower than that in local ULIRGs. If the
majority of high-redshift galaxies have lower levels of obscuration than Arp 299
(as indicated by the ratio of total IR to 8$\mu$m luminosity), we should assume a
lower missing SN fraction at high redshift for that fraction of the galaxies. Using
these considerations, we derive three models for the expected missing fraction
of CCSNe as a function of redshift.

The missing fraction of SNe in high-redshift U/LIRGs is highly uncertain. For
the main-sequence (i.e., non-starburst) galaxies, little is known about the
extinction, although they have been found to be disk-like and not as compact as
their local U/LIRG counterparts (Kartaltepe et al. 2012). But, given that they are
luminous in the mid- and far-IR, they contain substantially more gas and
dust than local disk galaxies. It therefore seems likely that the
extinction experienced by SNe in these galaxies is higher on average than in
the local population of normal galaxies. Furthermore, Reddy et al. (2012) found that roughly 80\%
of the star formation in typical star-forming galaxies at $z$ $\sim$ 2 is obscured
(at UV wavelengths) by dust. We assume that the level of obscuration in these
systems is roughly similar to that of the circumnuclear regions in Arp~299
and thus adopt a 37\% missing fraction of SNe in these galaxies.

To calculate the total missing fraction as a function of redshift, we use the relative
contributions to the cosmic star formation density of normal galaxies (defined
as galaxies with $L_{\rm IR}<10^{11} L_\odot$), LIRGs, and ULIRGs from Magnelli et al. (2011). We
then assume that SNe either explode in normal (i.e., with low-to-moderate IR luminosity)
galaxies, starburst U/LIRGs, or main-sequence (high-redshift) U/LIRGs. Following Magnelli et al.
(2011), the total SFR density at each redshift is defined as the sum of the unobscured SFR density
traced by the UV light (accounting for $\sim$20\% of the total SFR density) and obscured SFR density traced
by the IR light. We assume a zero missing
fraction for the SNe originating from the unobscured star formation. However, we note that our estimated
missing fraction in normal galaxies can include SNe originating from both obscured (e.g., SNe 2002hh in
NGC 6946) and unobscured star formation (e.g., SN 2008jb in a dwarf irregular host galaxy with a metallicity
similar to the Small Magellanic Cloud). We have therefore corrected this missing SN fraction to correspond
only to the obscured star formation in normal galaxies. For this correction we adopted a local fraction of
unobscured star formation in normal galaxies of 17\%, which we obtained by combining the unobscured UV star
formation from Schiminovich et al. (2005) with the IR star formation results of Magnelli et al. (2011) and
assuming the contribution of the unobscured UV light to the bolometric luminosity of U/LIRGs to be negligible.
Our estimated missing fractions in normal galaxies (both observed and corrected) and in U/LIRGs are summarized
in Table 8. In the following we adopt the results of Kartaltepe et al. (2012) according to which 42.6\% of the U/LIRGs at
$z$ $\sim$ 2 are starbursting. Since there is no strong redshift dependence in the Kartaltepe data at
1.5$<z<$2.5, we assume that this value is valid at $z$ = 1.5, and interpolate the fraction of starbursting
U/LIRGs between $z$ = 0 (100\%) and $z$ = 1.5 (42.6\%) assuming a simple linear evolution. At $z$ $>$ 1.5,
we keep the fraction constant.

The effects of host galaxy extinction on the detectability of SNe are also expected to
increase with redshift, since in general, shorter rest-frame wavelengths are observed at higher
redshifts. However, we note that current SN searches are often optimized for detecting SNe at
rest-frame $B$- and $V$-bands, since the SN SED peaks in this range. The uncertain nature of U/LIRGs
at high-$z$ currently far outweighs any wavelength-dependent effects on our predictions for the missing
SN fractions and we do not attempt to account for these. Therefore, our predictions are only valid
for rest-frame optical SN searches.

Bothwell et al. (2011) derived the local SFR distribution function and found $\sim$20\%
of the star formation occurring in starbursts defined as systems forming stars at $\geq$10 $M_{\odot}$~yr$^{-1}$.
They also studied the contribution of U/LIRGs to the local SFR volume density and found a
value of $\sim$10\%. Therefore, $\sim$10\% of the local SFR can be assumed to occur in starbursts
with $L_{\rm IR}<10^{11} L_\odot$ (e.g., M 82 and NGC 4038/9). However, the missing fraction of SNe in
such galaxies is not well constrained. In addition, we know little about their evolution as a function of redshift.
Given these substantial uncertainties we decide not to consider starburst galaxies with $L_{\rm IR}<10^{11} L_\odot$
separately and instead assume that these are included in our missing SN fraction estimated for the normal $L_{\rm IR}<10^{11} L_\odot$
galaxies. We expect that this decision does not significantly affect our final results. For example,
if adopting a similar missing SN fraction for the starburst galaxies as found for the circumnuclear
regions of Arp~299 instead of the value found for normal galaxies, we find that the local missing SN fraction
would increase only by two percentage points, which is negligible compared to the other uncertainties (see below).

Based on these assumptions, we derive three different models for the missing fraction and its dependence on redshift.
We call these {\it Nominal, Low}~and {\it High} missing fraction models (for summary see Table 9).

{\it Nominal model.} The nominal model is based on the assumption that the fraction of the U/LIRGs that are compact
and starbursting decreases with redshift, following the results from Kartaltepe et al. We
assume that Arp 299 represents local U/LIRGs. At high redshift we assume that Arp 299 only
represents the U/LIRGs that are starbursting, i.e., lie more than three times above the specific
star formation main-sequence locus. To calculate the missing fraction, we assume that all
starbursting U/LIRGs have a missing fraction of 83\%.
For the non starbursting U/LIRGs, we assume that the missing  fraction can be represented by
the value estimated for the circumnuclear regions of Arp 299, i.e., 37\% of missing SNe. For normal
galaxies, we assume a 19\% missing fraction.

{\it Low missing fraction.} Again, we assume that Arp 299 represents local U/LIRGs and at high redshift it only represents the
U/LIRGs that are starbursting. To calculate the missing fraction, we assume that all starbursting
U/LIRGs have a missing fraction of 83\%, while the $z$ $>$ 0 U/LIRGs that are not starbursting, have 
the same missing fraction as the normal galaxies (i.e. 19\%). For the low missing fraction case, we
further assume that the remaining normal galaxies have a missing fraction given by the 1$\sigma$ 
lower limit of our estimate, i.e., 7\%.

{\it High missing fraction.} We assume that all ULIRGs and LIRGs can be represented by Arp 299 at all
redshifts, having a missing SN fraction of 83\%. This should be an upper limit
since the SFR of Arp 299 is dominated by the compact and highly obscured
nuclei A and B1+B2 whereas the U/LIRG population at high redshifts is dominated by
galaxies forming stars in the ``normal'' main-sequence mode probably more
similar to the circumnuclear star formation in Arp 299. For this model, we also assume an upper 
limit for the missing fraction in normal galaxies by using the 1$\sigma$ upper limit, i.e., 43\% of the SNe missing.

\begin{table}
\caption{Missing SN Fraction}
\centering
\begin{tabular}{lccrcccc}
\hline
\hline
Redshift & Nominal & Low & High &$|$ &$f$(normal) & $f$(LIRG) & $f$(ULIRG)
\\
\tableline\tableline
  0.0 &0.189 &0.094 &~~0.381 &$|$& 0.146 & 0.040 & 0.003 \\
  0.1 &0.198 &0.108 &~~0.382 &$|$& 0.139 & 0.056 & 0.004\\
  0.2 &0.213 &0.127 &~~0.388 &$|$& 0.131 & 0.077 & 0.005\\
  0.3 &0.233 &0.152 &~~0.401 &$|$& 0.121 & 0.104 & 0.008\\
  0.4 &0.261 &0.184 &~~0.421 &$|$& 0.111 & 0.140 & 0.010\\
  0.5 &0.296 &0.223 &~~0.451 &$|$& 0.099 & 0.183 & 0.014\\
  0.6 &0.325 &0.252 &~~0.479 &$|$& 0.092 & 0.215 & 0.018\\
  0.7 &0.341 &0.266 &~~0.502 &$|$& 0.089 & 0.231 & 0.021\\
  0.8 &0.356 &0.279 &~~0.525 &$|$& 0.086 & 0.246 & 0.024\\
  0.9 &0.367 &0.288 &~~0.540 &$|$& 0.081 & 0.257 & 0.030\\
  1.0 &0.372 &0.293 &~~0.549 &$|$& 0.073 & 0.261 & 0.038\\
  1.1 &0.381 &0.300 &~~0.564 &$|$& 0.066 & 0.267 & 0.048\\
  1.2 &0.383 &0.299 &~~0.573 &$|$& 0.061 & 0.264 & 0.058\\
  1.3 &0.377 &0.291 &~~0.574 &$|$& 0.058 & 0.254 & 0.065\\
  1.4 &0.371 &0.282 &~~0.576 &$|$& 0.055 & 0.243 & 0.073\\
  1.5 &0.365 &0.273 &~~0.578 &$|$& 0.051 & 0.232 & 0.081\\
  1.6 &0.366 &0.275 &~~0.578 &$|$& 0.050 & 0.229 & 0.087\\
  1.7 &0.364 &0.273 &~~0.576 &$|$& 0.050 & 0.229 & 0.085\\
  1.8 &0.362 &0.272 &~~0.573 &$|$& 0.050 & 0.228 & 0.084\\
  1.9 &0.361 &0.271 &~~0.571 &$|$& 0.050 & 0.228 & 0.083\\
  2.0 &0.359 &0.269 &~~0.569 &$|$& 0.050 & 0.228 & 0.082\\
\tableline
\end{tabular}
\tablecomments{The missing SN fractions calculated for different
redshift bins following the assumptions
from Tables 8 and 9. Columns 5-7 show the contribution to the
nominal missing fraction from normal galaxies, LIRGs, and ULIRGs,
respectively.}
\end{table}

Using these assumptions and the evolution of the SFR and its contribution from
normal galaxies, LIRGs, and ULIRGs taken from Magnelli et al. (2011), we calculate how the missing
fraction of CCSNe evolves with redshift (see Table 10). In Figure 5, we plot the nominal model as a
solid line, while the {\it Low} and {\it High} missing fraction models are shown with dashed lines. We also compare these
with the prediction from Mannucci et al. (2007). The main differences between the two models is the higher
missing SN fraction locally in our case, which is a consequence of the relatively high fraction (2 out of 13)
of local CCSNe that showed a high amount of dust extinction even in moderately inclined host galaxies.
We also note that our missing SN fraction model levels out at $z\gsim1.2$, reflecting the fact that we
assume that high-redshift U/LIRGs are not as compact as the local ones, e.g., Arp 299, and therefore
have a lower fraction of missing CCSNe.

\begin{table}
\caption{Parameterization of the Missing SN Fraction}
\centering
\begin{tabular}{lcc}
\hline
\hline
Redshift & ~~~~~~~~~~$k$~~~~~~~~~~~ & ~~~~~~~~~~$m$~~~~~~~~~~\\
\tableline\tableline
$z<0.25$      & 0.135  & 0.188\\
$0.25<z<0.55$ & 0.314 & 0.144\\
$0.55<z<1.15$ & 0.116 & 0.253\\
$1.15<z<1.55$ & -0.049 & 0.442\\
$1.55<z<2.00$ & -0.016 & 0.392\\
\tableline
\end{tabular}
\tablecomments{A straight line parameterization for the nominal
missing SN fraction: $f$$_{\rm missing}$ =$k$$\times$$z$ + $m$.}
\end{table}

Our derived missing SN fraction can be well represented by straight lines within the different redshift
bins according to
\begin{equation}
f_{\rm missing}=k\times z + m
\end{equation}
where the values for [k,m] are listed in Table 11 for the nominal model. To correct the derived CCSN rates for the missing
fraction of SNe hidden in highly extinguished environments, the results have to be multiplied by a de-bias
factor given by
\begin{equation}
f_{\rm de-bias}=\frac{1}{1-f_{\rm missing}}
\end{equation}
\\
\\
\begin{figure}
\epsscale{1.0}
\plotone{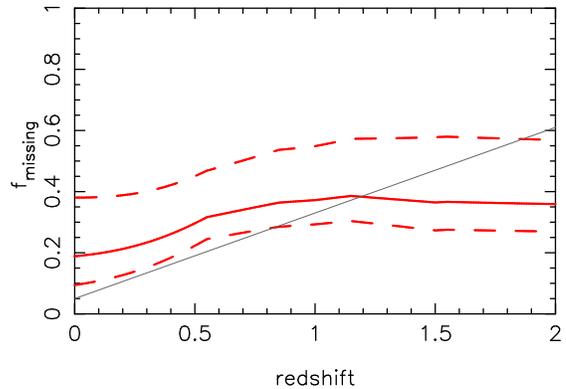}
\caption{Fraction of SNe missed by rest-frame optical searches as a function
of redshift. Red lines show our best ({\it Nominal}) estimate together with {\it Low} and
{\it High} missing fraction models as dashed lines. Solid black line is the missing
fraction from Mannucci et al. (2007).}
\label{figrate}
\end{figure}

\section{Discussion and Summary}
The SN events with high extinctions can have an important impact for SN statistics when estimating the
{\it complete} CCSN rates including the optically obscured SNe. This is essential
when using CCSNe as probes of the SFR at both low- and high-$z$ (e.g., Cappellaro et al.
1999; Dahlen et al. 2004, 2012; Botticella et al. 2008; Melinder et al. 2012) with
the aim of providing a new independent measurement of the cosmic star
formation history. Furthermore, accurate determination of the complete CCSN rates will
be crucial for comparison with the diffuse SN neutrino background in the future (e.g.,
Lien et al. 2010). We have shown that a substantial fraction of CCSNe have
remained undetected by current optical SN searches due to obscuration by large amounts
of dust in their host galaxies. We find that there should be missing CCSNe in highly extinguished 
environments in both normal host galaxies (even with moderate inclination) by $\sim$5-36\% and in
highly dust-enshrouded environments in U/LIRGs by up to $\sim$70-90\%. We note that our
estimated missing SN fraction in normal galaxies is also consistent with the recent findings
of Micha{\l}owski et al. (2012) for their sample of z $\lesssim$ 1 gamma-ray burst host galaxies.
For a volume-limited rest-frame optical SN survey we find the missing SN fraction to increase from its average local value
of $\sim$19\% to $\sim$38\% at $z$ $\sim$ 1.2 and then stay roughly constant up to $z$ = 2.

Using a local sample of CCSNe discovered during the last 12 yr, we find a lower limit for the local CCSN rate
of 1.5$^{+0.4}_{-0.3}\times 10^{-4}$ yr$^{-1}$ Mpc$^{-3}$
within the 6-15 Mpc volume, which is consistent with the CCSN rate estimate within 11 Mpc from Botticella et al (2012). 
Our estimated CCSN rate is significantly higher than the volumetric CCSN rate from the LOSS (Li et al. 2011a) of 0.84
$\pm$ 0.18 $\times 10^{-4}$ yr$^{-1}$ Mpc$^{-3}$ (scaled to correspond to $H_{0}$ = 70 km s$^{-1}$ Mpc$^{-1}$). If
applying our newly derived missing SN fraction correction the volumetric rate of LOSS would become 1.04 $\pm$ 0.22
$\times 10^{-4}$~yr$^{-1}$~Mpc$^{-3}$ or 1.35 $\pm$ 0.29 $\times 10^{-4}$~yr$^{-1}$~Mpc$^{-3}$ if adopting the ``high missing
fraction'' model. Therefore, within the uncertainties in the missing SN fraction correction, this is consistent with
our CCSN rate estimate within 6-15 Mpc. We note that LOSS made use of their observed luminosity functions of SNe
to compensate for the effects of host galaxy extinction for their derived SN rates. Their results indicated
that the line-of-sight extinctions toward SNe in the highly inclined galaxies were not significantly higher
than in the less inclined systems. However, they noted that this could still be a consequence of a small
number of SNe. However, there are also a number of other possible contributing
factors to the difference between the CCSN rate estimates, including the possibility of LOSS missing a
larger fraction of the intrinsically faint events than our 15 Mpc sample. Also, we note that the volumetric
CCSN rate of LOSS has been obtained by multiplying their estimated rates in units of galaxy $K$-band
luminosities (SNuK) with the local $K$-band luminosity density introducing additional uncertainties.

Our CCSN rate within 6-15 Mpc may also be elevated by cosmic variance. Assuming a Salpeter IMF between
0.1 and 125 $M_{\odot}$ and CCSN progenitor masses between 8 and 50 $M_{\odot}$, we find that our CCSN rate
corresponds to an SFR 0.021$^{+0.006}_{-0.005}$ $M_{\odot}$ yr$^{-1}$ Mpc$^{-3}$.
This is very similar to the local rate
0.019 $M_{\odot}$ yr$^{-1}$ Mpc$^{-3}$ of Horiuchi et al. (2011), where the latter rate is given for the
cosmology adopted here. The SFR within 11 Mpc based on the HUGS program (Kennicutt et al. 2008;
Bothwell et al. 2011) should, however, be similarly affected by cosmic variance, allowing a more direct comparison
between rates. The SFR derived in Bothwell et al. 0.023$^{+0.002}_{-0.002}$ $M_{\odot}$ yr$^{-1}$ Mpc$^{-3}$~
(given for our adopted cosmology) is consistent with our rate within the error bars. This is also similar to the rate of
Magnelli et al. (2009). We therefore conclude that our rate is consistent with what is expected from the SFR and there
is no need to correct for SNe missed in our nearby sample. Horiuchi et al. (2011) and Melinder et al. (2012) discuss
the choice of the IMF, and they show that the IMF dependence is mostly canceled out as long as the same IMF is used when originally
scaling from the massive star SFR to the total SFR and when converting between the CCSN rate and the SFR. Here we have adopted
the Salpeter IMF, which has also been used for deriving the SFRs. We also note that calculating the CCSN rate in the very
nearby universe within a distance of $<6$~Mpc, we find a significant increase in the rate by a factor $\sim$5 compared to
the rate found within 6-15 Mpc, caused by a few galaxies that have each hosted multiple SNe. This further supports the suggestion
of a significant local overdensity in the SFR within $\sim$10 Mpc (e.g., Karachentsev et al. 2004).

Horiuchi et al. (2011) note that local and low-redshift CCSN rates published before 2011, are lower compared to those
expected from the SFR by a factor $\sim$2 at a 2$\sigma$~confidence. As a solution to this discrepancy,
they suggested that there could be a population of faint CCSNe ($M \sim -15$) that are typically missed by SN surveys or
that there could be a population of SNe hidden by dust, or a combination of these two effects. Using their sample of CCSNe
a distance of within 11 Mpc, Botticella et al. (2012) did a detailed comparison with the local SFRs and found their CCSN rate
to be consistent with that expected from the SFR derived from far-UV luminosities and higher than expected based on H$\alpha$
luminosities. In our CCSN sample within 12 Mpc, there are no CCSNe fainter than $M$ $\sim$ -15 and roughly 20\% fainter than $M$ $\sim$ -15.5.
These intrinsically faint events are more likely to be missed in SN searches over a larger volume compared to our $<12$~Mpc sample and could
therefore lead to underestimates of the CCSN rate. Even if such events at the peak would be above the magnitude limit
of the survey, the time on the light curve they spend above the limiting magnitude of the search is shorter for this
population and unless accounted for, will lead to an underestimate of the rates.

Using our local sample of CCSNe, for which we were able to include SNe both with high extinctions and faint
intrinsic magnitudes, we do not find any discrepancy with the expectations from the SFRs,
even when taking the cosmic variance into account. Looking at the fraction of SNe missed in highly extinguished
environments, we have found locally $f_{\rm missing}$=19$^{+19}_{-10}$\%, corresponding to a de-bias factor of $\sim$1.1-1.6.
This is smaller than the factor $\sim$2 suggested by Horiuchi et al. (2011) but together with a realistic
fraction of intrinsically faint events can account for the apparent discrepancy between some of the previous local CCSN rate
estimates and the expectations from the SFRs. The effects of extinction correction on the CCSN rates at higher redshifts
are presented and discussed thoroughly in Melinder et al. (2012) and Dahlen et al. (2012). We therefore conclude
that correcting for the CCSNe missed due to very high dust extinctions in their host galaxies is crucial for deriving
accurate CCSN rates. Taking these effects into account should lead to CCSN rates that are consistent with
those expected from the SFRs.

\acknowledgments
We thank the anonymous referee for several useful suggestions, Enrico Cappellaro and Maria Teresa Botticella for
comments in the manuscript, and Nancy Elias de la Rosa and Chris Dudley for helpful discussions. We thank Benjamin Magnelli for providing
data over the redshift evolution of the IR luminosity densities.
S.M. and E.K. acknowledge financial support from the Academy of Finland (project: 8120503).
A.A.H.  thanks the Spanish Plan Nacional through grant AYA2010-21161-C02-01 and the Universidad
de Cantabria through the Augusto G. Linares program for financial support.
M.A.P.T. acknowledges support by the Spanish MICINN through grant
AYA2009-13036-CO2-01, partially funded by FEDER funds. This research
has also been partially funded by the Autonomic Government of
Andalusia under grants P08-TIC-4075 and TIC-126.
Based on observations obtained at the Gemini Observatory, which is operated by the
Association of Universities for Research in Astronomy, Inc., under a cooperative agreement
with the NSF on behalf of the Gemini partnership: the National Science Foundation (United
States), the Science and Technology Facilities Council (United Kingdom), the
National Research Council (Canada), CONICYT (Chile), the Australian Research Council
(Australia), Ministério da Ciência, Tecnologia e Inovação (Brazil) 
and Ministerio de Ciencia, Tecnología e Innovación Productiva  (Argentina).
We acknowledge the usage of the HyperLeda database (http://leda.univ-lyon1.fr).

\end{document}